\documentclass[journal,comsoc]{IEEEtran}

\usepackage[T1]{fontenc}

\usepackage{amsmath}
\usepackage{amsfonts}
\usepackage{amstext}
\usepackage{epsfig}
\usepackage{amssymb}
\usepackage{cite}
\usepackage{hhline}
\usepackage{multirow}
\usepackage{framed}
\usepackage{xcolor}
\usepackage{lipsum}
\usepackage{enumerate}
\usepackage{color}
\usepackage{caption}
\usepackage{subcaption}
\usepackage{graphicx}
\usepackage{algorithm}
\usepackage{algpseudocode}
\usepackage{mathtools}
\usepackage{booktabs}
\usepackage{bm}
\usepackage{cite}
\usepackage{url}
\usepackage{makecell}
\usepackage[english]{babel}
\usepackage{graphicx}
\DeclareMathOperator*{\argmax}{argmax} 

\usepackage{cite}

 \usepackage{pgfplots}
\pgfplotsset{compat=newest}
\usepackage{tikz}
\usetikzlibrary{plotmarks}
\usetikzlibrary{arrows.meta}
\usepgfplotslibrary{patchplots}
\usepackage{grffile}

\usetikzlibrary{shapes,arrows}

\usetikzlibrary{intersections}

\ifCLASSINFOpdf
 
\newtheorem{example}{Illustrative Example}

\begin{document}
	\title{	Learning-based Handover in \\Mobile Millimeter-wave Networks}
	\author{\IEEEauthorblockN{Sara Khosravi,~\IEEEmembership{ Student Member,~IEEE,} Hossein S. Ghadikolaei,~\IEEEmembership{ Member,~IEEE,}  and \\Marina  Petrova,~\IEEEmembership{ Member,~IEEE}}\\	
		\thanks{S. Khosravi,
			 H. S.  Ghadikolaei and M. Petrova are with school of EECS, KTH Royal Institute of Technology, Stockholm, Sweden (emails:\{sarakhos, hshokri, petrovam\}@kth.se). The work of H. S. Ghadikolaei was partially supported by the Swedish Research Foundation under grant 2018-00820.}}
		
	\maketitle
\begin{abstract}
Millimeter-wave (mmWave) communication is considered as a key enabler of ultra-high data rates in the future cellular and wireless networks. The need for directional communication between base stations (BSs) and users in mmWave systems, that is achieved through  beamforming, increases the complexity of the channel estimation. Moreover, in order to provide better coverage, dense deployment of BSs is required which causes frequent handovers and increased association overhead.
In this paper, we present an approach that jointly addresses the beamforming and handover problems. Our solution entails  an efficient beamforming method with a few number of pilots and a learning-based handover method supporting mobile scenarios. We use reinforcement learning algorithm to learn the optimal choices of the backup BSs in different locations of a mobile user. We show that our method provides an almost constant rate and reliability in all locations of the user's trajectory with a small number of handovers.
Simulation results in an outdoor environment based on narrow band cluster mmWave channel modeling and real building map data show the superior performance of our proposed solution in achievable instantaneous rate and trajectory rate.

\end{abstract}


%
\IEEEpeerreviewmaketitle
\renewcommand{\figurename}{Fig.}
\newcommand{\Tx}{\mathrm{Tx}}
\newcommand{\Rx}{\mathrm{Rx}}
\newcommand{\Base}{\mathrm{Base}}
\newcommand{\maximize}{\mathrm{maximize}}
\newcommand{\SNR}{\mathrm{SNR}}
\newcommand{\R}{\mathrm{R}}
\newcommand{\PS}{\mathrm{PS}}
\newcommand{\LoS}{\mathrm{LoS}}
\newcommand{\NLoS}{\mathrm{NLoS}}
\newcommand{\C}{\mathrm{C}}
\newcommand{\BS}{\mathrm{BS}}
\newcommand{\UE}{\mathrm{UE}}
\newcommand{\T}{\mathrm{T}}
\newcommand{\Aging}{\mathrm{Aging}}
\newcommand{\D}{\mathrm{D}}
\newcommand{\DM}{\mathrm{DM}}
\newcommand{\HO}{\mathrm{HO}}
\newcommand{\PL}{\mathrm{PL}}
\def\TT{\mathsf{T}}
\def\HH{\mathsf{H}}
\newcommand{\Pilot}{\mathrm{Pilot}}

\section{Introduction}
\IEEEPARstart{T}{he} expected growth of the mobile traffic and high data rate demands has triggered the design of communication systems that operate in millimeter-wave (mmWave) bands \cite{rappaport2013broadband}.
The main advantage of moving to the mmWave spectrum is the availability of  huge bandwidth in comparison to the conventional sub-6 GHz spectrum. 
However, mmWave bands are severely affected by obstacles unlike  the sub-6 GHz bands for two main reasons. First, due to an order of magnitude smaller wavelength, the signals cannot diffract well against most common materials in urban environments, leading to severe penetration loss and blockage \cite{rappaport2013broadband}. Second, the need for using directional communication to compensate for the high propagation loss, increases the beam misalignment chance, especially in the presence of many obstacles when there is a need to frequently update the beamforming vectors \cite{andrews2016modeling,kutty2016beamforming,shokri2015millimeter}. Establishing and maintaining mmWave links are even more challenging in mobile environments where both the  users and obstacles are moving. In order to provide good coverage and improve the capacity, base station (BS) densities need to be significantly higher in mmWave network \cite{UDN,rappaport2013millimeter}. These bring new challenges when compared to the sub-6 GHz networks \cite{hanoverHet}. For example, providing a reliable connection through UE's trajectory while balancing the number of handovers is a key design challenge to be solved if the mmWave networks are to support Gbps data rate for mobile users. In order to address this challenge, an efficient beamforming method is required to enable low latency and signaling overhead. In this paper, we jointly address the handover and beamforming challenges. In the next subsection, we describe our main contributions.
\subsection{Our Contributions}
The main objective of this paper is to develop an efficient and lightweight joint handover and beamforming method that maintains a predefined level of throughput along the UE's trajectory. More specifically, we address the following questions.

\begin{itemize}
	\item {How can sparsity and correlation of valid paths between the BS and the user equipment (UE) be used to design  an efficient beamforming}? 
	We propose a beamforming algorithm based on constructing and maintaining a database of path skeletons, i.e., available paths between the BS and the UE. In our proposed algorithm, beam searching will only run through the path skeletons not through all the directions. Moreover, our algorithm tracks the correlation between path skeletons in different locations of the UE and queries a new path skeleton only when a significant change such as a sudden blockage has occurred.
	\item {How can a handover method be designed to provide a reliable connection in mobile scenarios?}
	We propose a learning-based handover method based on keeping an updated backup channel for the serving BS. We use reinforcement learning (RL) to optimize the list of the backup BSs. Our  method comprises two decision making phases. In the first phase, our algorithm makes a decision regarding pinging a good backup BS. We model this selection process as  an RL problem that takes  the mobility prediction information as input and returns the best candidate BSs as the backup for a specific location. In the second decision making phase, our algorithm uses the channel estimation results of the both serving and backup BSs and makes the handover execution decision. 
	\item \textit{What are the benefits of our proposed method?} We compare our proposed approach with two relevant baselines in the literature. We numerically compare the key performance indicators including number of handovers, connection reliability, instantaneous rate and trajectory rate  of our approach with the baselines. The results indicate that our proposed approach substantially outperforms the baselines in terms of number of handovers and more importantly connection reliability throughout the trajectory. 
	
	\item \textit{What is the performance of our proposed method in a realistic channel model?} We use ray tracing with real building data map as the input. We also add common blockages like human bodies and cars to the simulation area in order to evaluate the performance of our method in a more realistic environment. 
\end{itemize}

We conclude that our solution is a signaling-efficient and lightweight approach that properly design the beamforming and handover so as to maintain a predefined quality-of-service level for the mmWave users in a dynamic and non-stationary environment. 
 In the following, we will review state-of-the-art approaches for beamforming and handover in mmWave networks.

\subsection{Related work}

Most of the mmWave beamforming approaches are carried out the exhaustive beam-searching over a set of pre-defined beams to find the beam pairs between a BS and a UE with the optimal alignment \cite{5174147,6171799}. However, due to the high dimension of the beam-searching, these approaches increase the overhead significantly.  Even other approaches, such as sparsity-aware beamforming \cite{heath2016overview} or subspace estimation \cite{ghauch2016subspace} suffer from the overhead in mobile scenarios. 
The recent compressive-sensing based approaches \cite{marzi2016compressive,heath2016overview,Hassanieh} need logarithmic number of the measurements during the beam-searching phase. However, due to the need of an adopted phase-array antennas \cite{Hassanieh} or phase coherent measurements \cite{marzi2016compressive}, they may not work well with existing mmWave devices.

 The authors in \cite{sur2016beamspy,8057188} presented the beam-searching methods which used the sparsity and the correlation of spatial channel response of mmWave channels in adjacent locations. Despite of their promising results in terms of throughput, those methods are validated on stationary users \cite{sur2016beamspy} or increase the complexity of the beam-searching phase \cite{8057188} thereby, constraining their applicability in real mobile scenarios. However, in this work, inspired by the idea of path skeleton in \cite{sur2016beamspy}, we propose an efficient approach to maintain the path-skeleton and track the correlation of path skeletons during the beam-searching phase. The overhead of our approach reduces as the number of UEs grows large, making it very useful for massive wireless access scenarios.

When it comes to the design of efficient and robust handover algorithms, the related state of the works can be grouped in three categories: learning-based handover \cite{MDP,RL2,stevens2008mdp,zang2018managing,learninhRL,side1,side3learning}, side information \cite{side1,side3learning,scaling,sur2018towards,side2}, and  multi-connectivity 
\cite{dual2,multicon1,multicon2,multicon3}.

To make the optimal handover decision, leveraging machine learning as the main decision maker tool can be an effective approach. In \cite{MDP}, authors used Markov decision process (MDP) in order to maximize the throughput and the achieved rate. However, due to the computation complexity of solving MDP, it cannot be directly applied to dense networks. The authors in \cite{RL2} proposed a novel handover policy  based on the RL framework for the radio access network slicing.  References \cite{stevens2008mdp} and \cite{zang2018managing} proposed a learning algorithm to manage vertical handovers in heterogeneous networks. The authors in \cite{learninhRL} introduced an RL-based handover policy named \textit{SMART} to reduce the number of the handovers while keeping the UE's quality-of-service in the heterogeneous network. However, the aim of our proposed algorithm is to maximize the long-term rewards (trajectory rate) and provide a reliable connection which is a function of both handover and beamforming in every slot. To this end, we propose an efficient beamforming method with low signaling overhead and propose a handover algorithm which maximizes the long-term reward by using statistical information of the mobility class and blockage distribution. Our approach, despite most of the the existing handover methods based on instantaneous change gain, prevent the ping-pong effect and lower trajectory rate due to the long-term view.  Based on the simulation results, our approach outperforms \textit{SMART} method in terms of the connection reliability and the number of the handovers.

 The authors in \cite{side1,side3learning} used the camera information to estimate the location of different obstacles and presented a proactive learning-based handover policy. However, due to the high density and variety of the obstacles in the urban environment, estimating the location of all obstacles may increase the network overhead. Our proposed learning-based handover method does not need online tracking of the obstacles and with keeping the connection toward a backup BS makes the handover decision. 

Side information or context-aware aided approaches make use of the location of the user or the obstacles in order to make a handover decision. The work in \cite{scaling} showed the importance of the location information in scaling mmWave networks to the dense and dynamic environment.
Authors in \cite{sur2018towards} proposed a handover method which leverages  channel measurement of dominant line-of-sight (LoS) path of serving BSs in order to estimate the LoS path properties of other BSs toward the UE and then ranks BSs based on  predicted beam strength. However, this method may not be applicable in all scenarios because this method cannot estimate good non-LoS (NLoS) paths in a crowded environment. However, our proposed approach considers all available paths in the path skeleton set of serving BS and backup BS toward the UE during the channel estimation phase and use the RL in order to select the backup channel. 

 In the multi-connectivity methods, a UE maintains its connection to multiple BSs (either at the mmWave or sub-6 GHz bands). Simultaneous connection of a UE with multiple BSs is analyzed vastly in \cite{dual2,multicon1,multicon2,multicon3} as a solution to the link failure and the throughput degradation in a dynamic environment. However, power consumption, synchronization and the necessity of frequent tracking are main challenges of multi-connectivity methods.
 For example, although different multi-connectivity schemes proposed in \cite{dual2} may improve the session-level mmWave operation in a realistic environment, the presented schemes need  additional connection-probe procedures and knowledge of the mmWave system state which add the overhead to the network. Our proposed approach is based on keeping UE's connection toward a backup mmWave BS with low overhead during the channel estimation by sending pilot signals only through the path skeleton sets.

 \subsection{Organization and Notation}
The rest of the paper is organized as follows. We introduce our system model in Section \ref{model}. Then, we describe our beamforming method and handover algorithm in Section \ref{method}. We model the problem of choosing backup BS as an RL problem in Section \ref{learning}. We numerically evaluate our algorithm in Section \ref{simulation}. Finally, we conclude our work in Section \ref{conclusion}.

\textit{Notation:} Matrices, vectors and scalars are denoted by bold upper-case ($\mathbf{X}$), bold lower-case ($\mathbf{x}$) and non-bold ($x$) letters, respectively. The $\ell_{2}$-norm, transpose, and conjugate transpose  of a vector $\mathbf{x}$ (or a matrix $\mathbf{X}$ ) are $\|\mathbf{x}\|$, $\mathbf{x}^\T$, and $\mathbf{x}^\HH$, respectively. We define set $[M]=\{1,2,..,M\}$ for any integer $M$.

\section{System Model} \label{model}
In this section, we introduce our main assumptions and system model. Table \ref{table1} summarizes our main notations.

 \begin{table}[t]
	\begin{center}
		\caption{Nomenclatures. }
		\label{table1}
		\begin{tabular}{|c|c|} 
			\hline
			\textbf{Notation} & \textbf{Description} \\
			\hline
			\hline 
					$j, N$ & Index and total number of $\BS$s in a zone \\
			$i, M$&  Location index and the length of a trajectory\\
			$p, P$& Index and total number of path clusters in a $\PS$ set\\
			$\ell, L$& Index and total number of SNR levels\\
		\hline
		$\SNR$&Signal to noise ratio\\
		$\PS$&Path skeleton set\\
			$\mathbf{f}, \mathbf{w}$& Beamforming and combining vectors\\		
			$\sigma^2$ & Thermal noise power\\
			$W$ & Signal bandwidth  \\
			\hline 
			$\mathbf{H}$&Channel matrix\\
			$\mathbf{u}_{\BS}(.), \mathbf{u}_{\UE}(.)$&Array response of BS and UE antennas\\
			$\theta^{\UE}, \phi^{\UE}$ & Horizontal and vertical AoA \\
			$\theta^{\BS}, \phi^{\BS}$ & Horizontal and vertical AoD \\
			$h_{rp} $& channel gain of $r$-th subpath of path cluster $p$\\
			$N_{\BS}, N_{\UE}$ & Number of $\BS$ and UE antennas\\
			

	\hline 
	
				\multicolumn{2}{|c|}{Handover algorithm parameters
		} \\
	\hline
		$\BS^{k}_{S}, \BS^{k}_{B}$&Serving and backup BS in CI $k$\\
		 $\SNR^\ell(\BS^{k}_{S},i)$&SNR level $\ell$ from $\BS^{k}_{S}$ toward UE in location $i$ \\	
		 $t_{\text{Log}}$& Age of the SNR in terms of CI\\

				\hline
		$\T_{\HO}$& Handover threshold \\
		$\T_{\D}$&PS distance threshold\\
		$\T_{\Aging}$&Aging threshold of a PS in the database\\
		\hline 
	\end{tabular}
\end{center}
\end{table}

We consider the downlink of a mmWave network with $N$ BSs and mobile UEs. We assume a two-dimensional Poisson point process (PPP) with density $\rho$  for the spatial distributions of the BSs, though our proposed algorithmic framework can work for any other model. We assume all BSs allocate equal resources to their serving UEs. Extension to the load balancing at the BS for multiple UEs scenario are left for the future work.

We employ a narrow band cluster 3D channel model \cite{andrews2016modeling}  with small number of clusters and $N_{\BS}$ antennas at the BS and $N_{\UE}$ antennas at the UE side. In this model the channel matrix $\mathbf{H}\in\mathbb{C}^{N_{\BS}\times N_{\UE}}$ between a UE in location $i$ of a trajectory (with $M$ points) and $\BS_j , ~j\in [N]$ is fixed during a coherence interval (CI) and can be defined as:
\begin{equation}
\mathbf{H}(j,i)
\!=\!{\frac{1}{\sqrt{R}}}\sum_{p=1}^{P}\sum_{r=1}^{R} h_{r,p} \mathbf{u}_{\text{\tiny UE}}(\theta^{\UE}_{r,p},\phi^{\UE}_{r,p}) \mathbf{u}^\HH_{\text{\tiny BS}}(\theta^{\BS}_{r,p},\phi^{\BS}_{r,p}),
\label{H}
\end{equation}
where $P$ is the number of path clusters and $R$ is the number of subpaths in each cluster. Each subpath has the horizontal and vertical angle of arrivals (AoAs), $\theta^{\UE}_{r,p},\phi^{\UE}_{r,p}$, and horizontal and vertical angle of departures (AoDs), $\theta^{\BS}_{r,p},\phi^{\BS}_{r,p}$, respectively. $h_{rp}$ is the complex gain of $r$-th subpath of cluster $p$ which includes both the path loss and small scale fading \cite{andrews2016modeling}. We generate these parameters based on different distribution as given in \cite[Table I]{andrews2016modeling}. For the sake of notation simplicity, we drop the notation $i$ and $j$ from the the channel parameters, whenever they are clear from the context. We consider a half wavelength uniform planar arrays of antennas both at the BS and the UE sides which can be defined as
 \cite{hemadeh2017millimeter}:
\begin{equation}
\mathbf{u}_{s}(\theta^s,\phi^s)=[1,..., e^{j\pi[n_\BS \sin(\theta)\cos (\phi)+n_\UE \sin(\theta)\sin(\phi)]},... ]^T
\end{equation}
where $1\leq n_\BS\leq N_\BS-1$, $1\leq n_\UE\leq  N_\UE-1$, and $s\in\{\UE,\BS\}$.

We use the following probability functions obtained based on the New York City measurements in \cite{samimi2015probabilistic} to define the probability of LoS and NLoS states of each link:
\begin{subequations}
	\begin{align}
	&p_{\LoS}(d)=\left[\min \left(\frac{27}{d},1\right).\left(1-e^{-\frac{d}{71}}\right)+e^{-\frac{d}{71}}\right]^2
	\\& p_{\NLoS}(d)=1-p_{\LoS}(d),
	\end{align}	
\end{subequations}
where $d$ is the 3D distance between UE and BS in meters. We model the pathloss of the LoS and NLoS links as:
\begin{equation}
\PL(d)[\text{dB}]=10 \log_{10}\left(\frac{4\pi d_{0}}{\lambda}\right)^2+10 \hat{n}  \log_{10}\left(\frac{d}{d_{0}}\right)+X_{\mu},
\end{equation}
where $d_0$ is the close-in free space reference distance which in this work $d_0=1$, $\lambda$ is the wavelength, $\hat{n} $ is the path loss exponent , and $X_{\mu}$ is a zero mean Gaussian random variable with the standard deviation $\mu$ in dB which represents the shadow fading. $\hat{n}$ and $\mu$ have different amounts for LoS and NLoS links. These parameters are given in \cite [Table V and VI]{rappaport2015wideband}.

The signal to noise ratio (SNR) at the UE in location $i$ which is serving by $\BS_j$ can be defined as

\begin{equation}
\SNR(j,i)=\dfrac{|\mathbf{w}^\HH(i)\mathbf{H}(j,i)\mathbf{f}(j,i)|^2}{\sigma^{2}W},
\end{equation} 
 where $\mathbf{f}\in \mathbb{C}^{N_\BS}$ is the beamforming vector in the BS side and $\mathbf{w}\in \mathbb{C}^{N_\UE}$ is the combining vector in the UE side, $W$ is the system bandwidth, $\sigma^{2}$ is the noise power level which is normalized by the transmit power. Due to the noise dominant nature of highly directional mmWave transmission \cite{andrews2016modeling}, we omit the interference effect of other BSs. We define the achievable rate per second, between BS $j$ and UE in location $i$ as $\R(j,i)=W\log(1+\SNR(j,i))$.
 
\begin{figure}[t]
	\centering
	\includegraphics[width=0.55\textwidth]{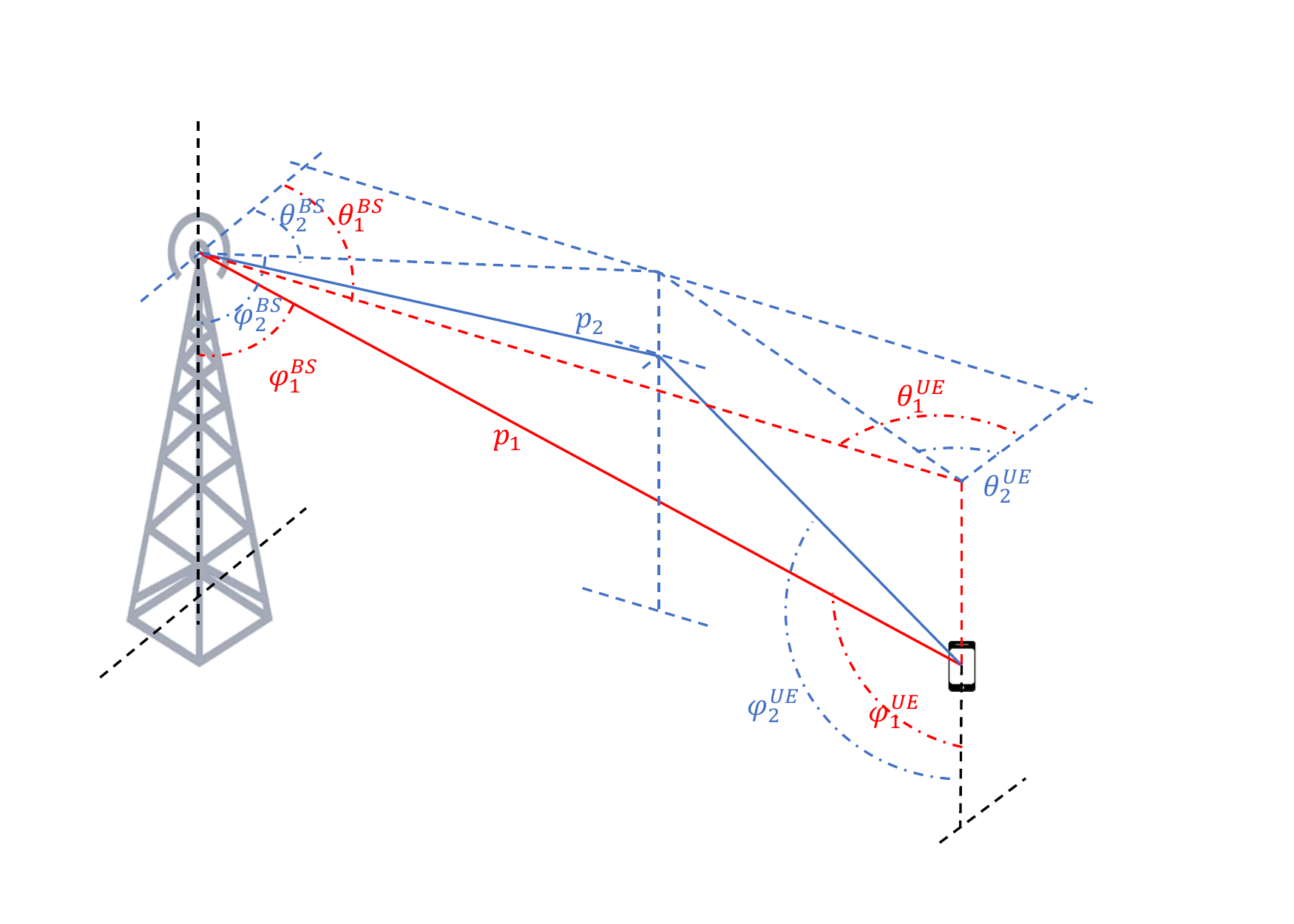}
	\caption{An example of a path skeleton set contains a LoS path ($p_1$) and a NLoS path ($p_2$)}.
	\label{fig1}
\end{figure}

 In order to find the available path clusters between the $\BS_j$ and the UE in location $i$, we define a set of path skeletons, $\PS(j,i)=\{p_{1},...,p_{P}\}$ with size $P$.
 Each path $p$ is defined based on AoA ($\theta^{\UE}_{r,p},\phi^{\UE}_{r,p}$), AoD ($\theta^{\BS}_{r,p},\phi^{\BS}_{r,p}$,) and the channel \mbox{gain ($\tilde{g}=\sqrt{\PL(d)}$)}. Due to the sparsity of the mmWave channels in angular domain, $P$ is a small number. Moreover, path skeleton sets have a correlation in adjacent locations \cite{sur2016beamspy}. A path skeleton can be identified using an exhaustive search method \cite{6171799} or the proposed method in \cite{sur2016beamspy}.
 An example of a path skeleton, $\PS=\{p_{1},p_{2}\}$, between a BS and a UE is illustrated in Fig. \ref{fig1}. For the sake of simplicity, one subpath of each path cluster is shown in this figure.

  We define a zone based on a certain geographical area. We consider one agent for each zone that connects to all the BSs in its zone. Fig. \ref{fig:fig1} shows our simulation area and an example zone with five BSs.
  We consider the pedestrian and the vehicular mobility models, modeled through some trajectories. The different trajectories are shown in Fig. \ref{fig:fig1}. 

\section{Proposed Method}
\label{method}

During the mmWave handover process, there are multiple potential $\BS$s to which the UE can connect. In mobile mmWave networks, the channel quality may drop quickly because of the mobility of the UE and temporary blockage by obstacles. Hence, re-execution of the beam-searching process to find a new serving $\BS$ may increase overhead and adversely impact on the UE throughput. In other words, beam searching with all the $\BS$s in the UE's vicinity, increases the complexity of the handover process especially in moderate to high $\BS$ densities.

In order to address the aforementioned challenges, we propose to maintain an ordered list of backup $\BS$s. In our proposed method, in case the link toward the serving $\BS$ dropped below a certain quality, a new link is established from the backup $\BS$ to the user with no need to search over all $\BS$ in the zone.

Our proposed handover mechanism consists of three components: pilot design and channel estimation, mobility prediction, and handover algorithm. 
We first design the pilot signals and then estimate the channel toward the serving BS using those pilots while assuming the mobility model of UE is available. Next, the handover execution decision is made based on the proposed handover algorithm. Our novel approach uses RL to select the backup channels and acquires the channel toward one non-serving $\BS$ at every coherent interval (CI). Using statistical data, we optimize this back-up inquiry process.
  In the following, we illustrate in detail these components. 

\tikzset{%
	bodyy/.style={inner sep=0pt,outer sep=0pt,shape=rectangle,draw,thick},
	dimen/.style={<->,>=latex,thin,every rectangle node/.style={fill=white,midway,font=\sffamily}},
	dimen1/.style={->,>=latex,thin,every rectangle node/.style={fill=white,midway,font=\sffamily}},
	symmetry/.style={dashed,thin},
}

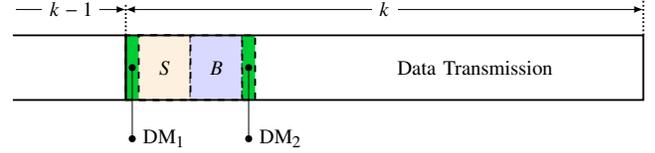
\begin{figure}[t]
	\centering
	\scalebox{0.86}{\hspace{-3mm}\small{\begin{tikzpicture}

    \node [bodyy,dashed,minimum height=1cm,minimum width=8mm,anchor=south west, fill=blue!20!green!] (C1) at (2,0) {};
    \node [bodyy,dashed,minimum height=1cm,minimum width=8mm,anchor=south west,fill=yellow!50!red!12] (S) at (2.2,0) {$S$};
    \node [bodyy,dashed,minimum height=1cm,minimum width=8mm,anchor=south west, fill=blue!15!white!] (B) at (3,0) {$B$};
    \node [bodyy,dashed,minimum height=1cm,minimum width=2mm,anchor=south west, fill=blue!20!green!] (C2) at (3.8,0) {};
    \node [bodyy,thick, minimum height=1cm,minimum width=2cm,anchor=south west] (body1) at (0,0) {};
\node [bodyy,draw=none,minimum height=12mm,minimum width=3mm,anchor=south west,fill=white]
(body3) at (-0.05,-0.1) {};
\node [bodyy,thick,minimum height=1cm,minimum width=8cm,anchor=south west] (body2) at (2,0) {};
\node [bodyy, draw=none, minimum height=1cm,minimum width=5.2cm,anchor=south west] at (4.8,0) {Data Transmission};
\draw[draw=none] (3mm,1) -- ++(0,+0.4) coordinate (D1) -- +(0,+3pt);
\draw[densely dotted,thick] (body1.north east) -- ++(0,+0.4) coordinate (D2) -- +(0,+3pt);
\draw [dimen1] (D1) -- (D2) node {$k-1$};
\draw[densely dotted,thick] (body2.north east) -- ++(0,0.4) coordinate (D3) -- +(0,4pt);
\draw [dimen] (D2) -- (D3) node {$k$};
\node[outer sep=0pt,circle, fill,inner sep=1.1pt] (C11) at (2.1,0.5) {};
\node[outer sep=0pt,circle, fill,inner sep=1.1pt, label={right:$\DM_{1}$}] (C12) at (2.1,-0.6) {};
\draw (C11) -- (C12);
\node[outer sep=0pt,circle, fill,inner sep=1.1pt] (C21) at (3.9,0.5) {};
\node[outer sep=0pt,circle, fill,inner sep=1.1pt, label={right:$\DM_{2}$}] (C22) at (3.9,-0.6) {};
\draw (C21) -- (C22);
\end{tikzpicture} }}
	\caption{Learning-based pilot design during CI $k$. Mini-slot $S$ and mini-slot $B$ are pilot transmission durations in order to estimate the channel toward serving $\BS$ and backup $\BS$, respectively. $\DM_{1}$ is the decision making phase regarding the choice of the backup $\BS$. $\DM_{2}$ is the decision making phase regarding the handover execution. }
	\label{fig2}
\end{figure}

\subsection{Channel Estimation} 
As shown in Fig. \ref{fig2} for a CI $k$, our proposed  pilot consists of a mini-slot $S$ to acquire the channel toward the serving $\BS$ channel and mini-slot  $B$ for estimating the channel toward a backup $\BS$. Each channel estimation mini-slot consists of $P$ pilots, where  $P$ is the size of the path skeletons. We consider two decision making ($\DM$) phases. In $\DM_{1}$, a backup BS for CI $k$ will be determined based on the optimal policy of the RL algorithm and in $\DM_{2}$, the decision regarding the execution of the handover will be made. We define an acceptable handover threshold ($\T_{\HO}$) based on the user quality-of-service. In other words, 
during the $\DM_{2}$ if the channel quality of the serving BS becomes lower than the predefined $\T_{\HO}$  for a certain duration (which can be defined based on UE's quality-of-service), UE will switch to the backup BS.   
In the following, we will summarize our proposed efficient beamforming method, which we proposed  in our recent work in \cite{sara}.

In the channel estimation phase, we consider a path skeleton database in each $\BS$ that contains the path skeletons of different locations in the coverage area of every BS. However, having a path skeleton database entails two cost terms: query and maintenance. Query cost refers to the limited budget that $\BS$ can query a new path skeleton from the database and maintenance cost is the cost of building and keeping the database updated. First, we focus on the query cost and assume that an updated path skeleton database is available for all $\BS$s. Then, we discuss the maintenance cost.

During the pilot transmission phase, the UE requests the path skeleton of its current location $(x_i,y_i)$ from the serving $\BS_j$. The pilot sequence is sent through the $P$ paths of $\PS(j,i)$ in order to estimate the channel between the UE and the $\BS$.
This estimation will then be used to design a precoding vector ($\mathbf{f}$) at the $\BS$ (from a given codebook $\mathcal{F}$) and a combining vector ($\mathbf{w}$) at the UE (from a given codebook $\mathcal{W}$) for the data transmission phase. Formally, we solve the following beamforming optimization problem: 
\tikzstyle{decision} = [diamond, draw, fill=blue!10, 
text width=4em, text centered, node distance=3cm, inner sep=0pt]
\tikzstyle{block} = [rectangle, draw, fill=white, 
text width=7em, text centered, rounded corners, minimum height=2em]
\tikzstyle{line} = [draw, -latex']
\tikzstyle{cloud} = [draw, ellipse,fill=red!10, node distance=3.5cm,
minimum height=2em]
\tikzstyle{io} = [rectangle, draw, fill=blue!5, 
text width=6em, text centered, minimum height=2em]

\begin{subequations}
	\begin{equation}
	\begin{aligned}
	& \underset{\mathbf{f}, \mathbf{w}}{\text{maximize}}
	&& |\mathbf{w}^\HH\mathbf{H}\mathbf{f}|^2 
	\end{aligned}
	\end{equation}
	\begin{equation}
	\begin{aligned}
	& \text{subject to}
	& \mathbf{f} \in \mathcal{F}, 
	\end{aligned}
	\end{equation}
	\begin{equation}
	\begin{aligned}
	&&&&&&&&&&\mathbf{w} \in \mathcal{W}.
	\end{aligned}
	\end{equation}
	\label{codebook}
\end{subequations}
In an environment with a small number of scatters, the optimal beamforming and combining may adjust to the array response of the strongest available path \cite{andrews2016modeling}. More details regarding the solution of \eqref{codebook} is provided in Appendix~A.

Due to the correlation of the path skeletons in adjacent locations \cite{sara}, there is no need to query a new path skeleton in every location of the UE. In other words, the $\BS$ can track the path skeleton changes and only ask a new beamforming solution when the current one is blocked or weakened by the obstacles. We consider the current path skeleton as the reference path skeleton  $\PS(j,0)$ that is known to both the UE and the $\BS_{j}$. In a new location $(x_i,y_j)$, the agent uses $\PS(j,0)$ to estimate $\PS(j,i)$ and $\mathbf{H}(j,i)$. We define the distance between the reference path skeleton and estimated path skeleton as a metric to assess the validity of using  the reference path skeleton in the new location $(x_i,y_i)$:
\begin{equation}
d(x_{i},y_{i} ; x_0,y_0) = \|\mathbf{\PS}(j,i)-\mathbf{\PS}(j,0)\|_{2}.
\end{equation}
Observations of \cite{sara} show that once the distance is sufficiently close (namely $d(x_{i},y_{i} ; x_0,y_0) \leq \T_{\D}$  for some small positive $\T_{\D}$),  the UE can use $\PS(j,0)$ to estimate the channel  $\mathbf{H}(j,i)$ in the new location. Otherwise, $\BS_{j}$ declares a significant change in the dominant paths. It then quires a new path skeleton and informs the UE. In this case, the reference path skeleton will be updated to the new path skeleton and $\BS_{j}$ tracks the validity of this new reference skeleton for beam-searching over time. We define $\T_{\D}$ as the decision threshold that highly depends on the network topology. Smaller $\T_{\D}$ results in frequent updates of the path skeletons and a higher overhead cost. Larger $\T_{\D}$ reduces the network overhead but may result in sub-optimal selected beamforming and combining directions through the UE trajectory. 
\begin{figure}[!t]
	\begin{tikzpicture}[node distance = 1.5cm, auto]
	\tikzstyle{every node}=[font=\footnotesize]
	\node [block] (init) {Extract a grid ID from the watch list};
	\node [cloud, right of=init] (system) {Database};
	\node [block, below of=init,yshift=0.3cm ] (rx) {Detect a UE in the grid ID};
	\node [decision,below of=rx,yshift=1cm] (decide1) {Does the UE confirm?};
	\node [block, left of=decide1, xshift=-1.8cm ] (rx2) {Choose another UE};
	\node [block, below of=decide1, yshift=-0.5cm] (ps) {Start the path skeleton process};
	\node [block, below of=ps] (store) {Store the path skeleton with the grid ID in the normal list};
	\node [decision, below of=store,yshift=0.6cm] (decide2) {Aging counter$ >\T_{\text{Aging}}$ ?};
	\node [io, left of=decide2, xshift=-1.8cm ] (store2) {The path skeleton is ready to use};
	\node [block, below of=decide2, yshift=-1cm] (remove) {Remove the path skeleton from the normal list and add its grid ID to the watch list};
	
	\path [line] (init) -- (rx);
	\path [line] (rx) -- (decide1);
	\path [line] (decide1) -- node {yes}(ps);
	\path [line] (decide1) -- node {no} (rx2);
	\path [line] (rx2) |-  (rx);
	\path [line,dashed] (system) -- (init);
	\path [line] (ps) -- (store);
	\path [line] (store) -- (decide2);
	\path [line,dashed] (store) -| (system);
	
	\path [line] (decide2) -- node {yes}(remove);
	\path [line] (decide2) -- node {no} (store2);
	\path [line,dashed] (remove) - | (system);
	\end{tikzpicture}
	\caption{The process of building and updating the database. First, a gridID is extracted from the watch list. If a detected UE in the grid ID approves the path skeleton finder request, the process starts and the skeleton set stores in the normal list with active aging counter. If the value of the aging counter exceeds the aging threshold, the skeleton set is removed from the normal list and its grid ID is added to the watch list. In this case, the algorithm returns to the skeleton finder loop \cite{sara}.}
	\label{flochart}
\end{figure}
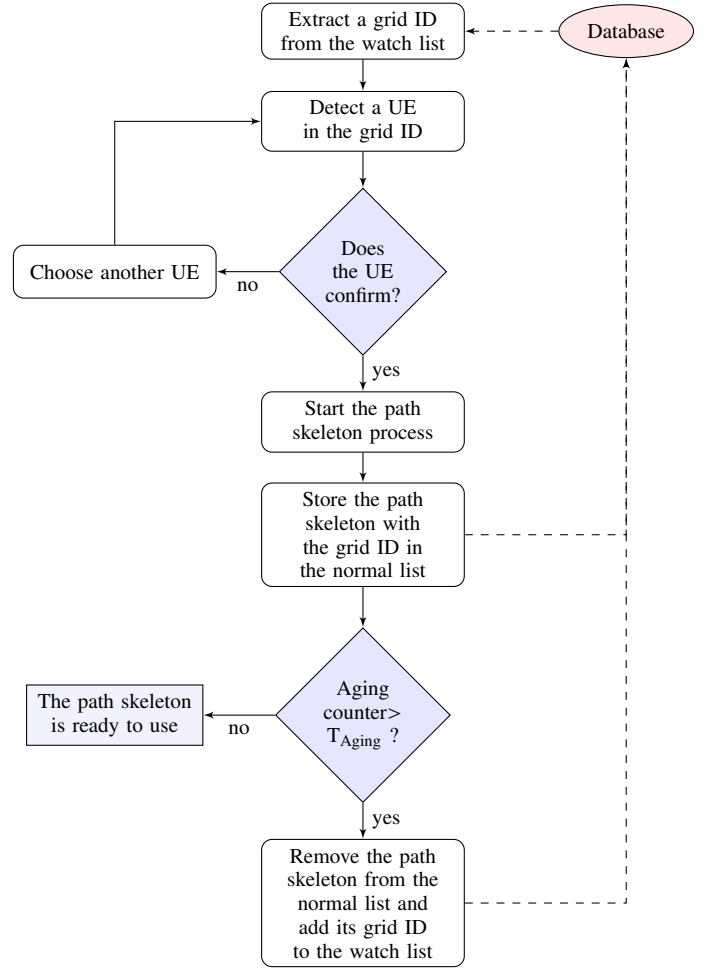
In order to choose the optimal $\T_{\D}$ for a limited query budget ($U_{\max}$), we run our algorithm for different  mobility models and trajectories in the coverage area of all $\BS$s. For instance, in the coverage area of $\BS_{j}$, the optimal threshold $\T_{\D}^\star(j)$ for pedestrian mobility model is the solution of the following optimization problem: 

	\begin{subequations}
		\begin{equation}
		\begin{aligned}
		& \T_{\D}^\star=\argmax_{\T_{\D}>0}  
		\sum_{i \in [M]} \mathbb{E}[\R(j,i)]
		\end{aligned}
		\end{equation}
		\begin{equation}
		\begin{aligned}
		& \text{subject to}
		&& \Pr\{U > U_{\max}\} \leq \delta,
		\end{aligned}
		\end{equation}
		\label{T}
	\end{subequations}
\!\!where $\R(j,i)$ is the rate between the UE in location $i$ and $\BS_j$, $U$ is the number of times the PS renewed, and  $\delta$ is an small given parameter. We have chosen $\delta=0.2$ in our framework.
In order to solve the optimization problem, we have used a well-known golden-section search method \cite[Section 7.2]{Chong} through a dataset of different pedestrian trajectories with length $M$.
 Note that if the estimated value of $\Pr\{U > U_{\max}\}$ is greater than $\delta$, we assume the corresponding value of the objective function is $-\infty$.
Note that these processes are done offline, so the complexity does not cause delay to the real time system.

The efficiency of our proposed beamforming method can be defined based on three parameters: computational and signaling complexity, throughput efficiency and energy consumption. In terms of throughput, our approach guarantees a close-to-optimal performance by updating the beamforming and combining vectors. Moreover, by sending pilots only over the path skeletons, our approach substantially reduces the beam searching overhead (the number of beams required to find the best alignment), making it efficient  in terms of energy consumption, computational and signaling complexity. 

Now, we present our approach regarding the building and maintaining the path skeleton database. In this case, we assume that every $\BS$ divides its coverage area to small grids and assigns a unique ID to them. The size of all the grids is equal and is chosen based on the network topology and balancing the complexity of building a database. Each grid is approximated to one point and  one path skeleton is recorded for each grid ID in the database. In other words, only one path skeleton finding process (like the one in \cite{sur2016beamspy} or exhaustive beam-searching \cite{6171799}) will run for each grid in order to build the path skeleton.

The path skeleton database has the normal list and the watch list that can be defined as the list of grid IDs with updated path skeletons and a list of grid IDs whose path skeletons are needed to be updated, respectively.
As it is shown in Fig. \ref{flochart}, a BS sends the path skeleton finder request to the UEs in every grid ID. It is important to note that a UE may refuse the request due to for example low battery level. If a UE accepts the request, the skeleton finder process will start and the skeleton is recorded in the normal list with a specific aging counter. If the aging counter of grid IDs in the normal list exceeds the predefined threshold ($\T_{\Aging}$), the BS will remove them from the normal list and adds them to the watch list. Fig.  \ref{flochart} illustrates this process.

$\T_{\Aging}$ depends heavily on the network topology. In a crowded urban environment, the channel conditions will change  rapidly so the database may need frequent updates. It means that  $\T_{\Aging}$ should be shorter for a highly dynamic environment compared to a stationary environment.

The overhead of the building and updating the database can be defined as the number of path skeleton finder requests that a UE will receive through its trajectory. In the crowded urban environment that the number of UEs is high, the database overhead is divided between all the users. Hence, the database overheat is almost negligible. More details of the proposed algorithm and performance evaluations are available in our recent work in \cite{sara}.

Notice that there are two overhead terms involved with path skeleton. A) path skeleton database, which maintains the most updated path skeletons and B) beam-searching over an existing path skeleton. Upon handover, we only have the overhead B, whose complexity scales with the path sets in the skeleton. This cardinality is very small due to the sparse scattering nature of a mmWave channel. The main overhead comes from A, which we continuously run in background, even when no handover is requested. In particular, the BS sends the path skeleton finder request to the UEs in different grids. If the UE accepts this request, this process will be started. 	
In the case that a grid does not have any UEs or all the UEs in that grid reject the path skeleton finder request, BS will start the path skeleton building process when a UE in that grid aims to establish a new connection or a handover. In other words, in the worst case, our method works like the existing methods for beam-searching based on path skeleton. However, such extreme case may rarely happen in a realistic urban environment. 
For example, assume that there exists $U$ UEs in each grid ID at each time slot and each one of them accepts the path skeleton building request with probability $p$, independent to the other UEs \footnote{This simple model, while not being realistic, gives the insights.}. Hence, the probability that no UE accepts the request during $\text{T}_{\text{Aging}}$ slots decreases with $U$ and $\text{T}_{\text{Aging}}$, i.e.,  $(1-p)^{U\text{T}_{\text{Aging}}} $.

\subsection{Mobility Prediction}
The mobility behavior and localization of the UEs are the important challenges in communication networks due to the key applications in handover and resource management \cite{mobility3}. An accurate prediction of the mobility pattern in dense mobile networks reduces the signaling overhead during the handover process \cite{mobility2,mobility5} and provides better quality-of-service and continuous connections.

 As it is shown in \cite{mobility1} the mobility behavior of UEs in mmWave networks is predictable with good performance. Most of the mobility patterns like pedestrians  or vehicles are destination or direction oriented. 
The different mobility prediction schemes existing in the literature use Markov chain, hidden Markov model, artificial neural networks, Bayesian network and data mining (for more details see \cite{mobility1}). The outputs of these prediction methods are the UE's moving direction, transition probability to the next location, the future location and the trajectory \cite{mobility1,mobility4}.
Studies in \cite{mobility11,scaling} show that due to the use  of the massive antenna arrays and the presence of the multi-path channel in mmWave networks, the performance of UE's localization is sufficiently high  in both uplink and downlink communication.
In this work, we assume that the mobility prediction information includes the UE's current location and its trajectory are available. The agent provides the mobility information to all $\BS$s in its zone.

\subsection{Handover Algorithm}
As mentioned in the previous subsection, the mobility prediction information, including UE's trajectory with length $M$ and its current location $i\in [M]$, is the input to our proposed handover algorithm. We assume one agent for a zone with $N$ BSs where $j\in [N]$ is the index of $\BS$s.

  We quantize $\SNR(j,i)$ to $L$ levels. During the channel estimation in both mini-slots $S$ and $B$, we only report  the SNR level $\SNR^{\ell, p^{\ast}}(j,i)$, $ \ell \in [L]$, where $p^{\ast}$ is the strongest path cluster over $\PS(j,i)$, $p^{\ast}=\argmax_{p} h_{r,p},~ p\in \PS(j,i)$. For the sake of notation simplicity, we drop the superscript of $p^{\ast}$ and write $\SNR^{\ell}(j,i)$. We define the handover threshold ($\T_{\HO}$) as the minimum acceptable $\SNR$ level. This parameter is determined based on the target quality-of-service level of the UE. 
  
 For the tagged UE, $t_{\text{Log}}$ denotes the age of the current $\SNR$ log toward $\BS_{j}$ in term of the number of CIs. The initial value of the $t_{\text{Log}}$ for all $BS_{j} , j\in[N]$ is equal to $t_{\text{Log}}=+\infty$.
 Once $\SNR^{\ell}(j,i)$ is obtained for any $i\in [M]$, $t_{\text{Log}}(BS_{j})$  is set to zero. 
 At the end of each CI $k$, $k\in\{1,2,...\}$, 
  $t_{\text{Log}}(BS_{j}), \forall j\in [N]$ is increased by $1$ to indicate the age of the current log. 

As it is illustrated in Algorithm \ref{Alg1}, in location $i$ and \mbox{CI $k$} channel estimation toward the serving BS ($\BS^{k}_{S}$) and backup BS ($\BS^{k}_{B}$) starts during the mini-slot $S$ and mini-slot $B$, respectively. If the SNR level of $\BS^{k}_{S}$ remains above  $\T_{\HO}$, the serving BS in the next CI will not be changed; otherwise, the handover decision will be made during $\DM_{2}$. If the $\SNR$ level of $\BS^{k}_{S}$ drops lower than  $\T_{\HO}$, for a certain time interval, the handover will be triggered. Then, a BS with an acceptable $\SNR$ (larger than $\T_{\HO}$) and a minimum amount of $t_{\text{Log}}$ (the most recent $\SNR$ updated)  will be selected as the main candidate for the handover.
\begin{example}
Consider  a zone with four $\BS$s. The $\SNR$ is quantized to two levels $\{\ell_{1},\ell_{2}\}$ and $\T_{\HO}$ is equal to $\ell_{1}$. Assume that in CI $3$, the $\BS^{3}_{S} = \BS_{1}$ and  $\BS^{3}_{B} = \BS_{4}$. The channel estimation will be done during mini-slots $S$ and $B$ and the SNR level and $t_{\text{Log}}$ of $\BS_{1}$ and $\BS_{4}$  will be updated. Now in $\DM_{2}$ the decision regarding the handover execution will be made. As it is shown in the following table, the $\SNR$ level of $\BS_{1}$ and $\BS_{4}$ are lower than $\T_{\HO}$, so the handover will run and the decision is the most recent updated BS with $\SNR$ level equal to $\ell_{2}$. In this case between  $\BS_{2}$ and $\BS_{3}$, the $\BS_{2}$ is selected because $t_{\text{Log}}(\BS_{2})<t_{\text{Log}}(\BS_{3})$. At the end of the CI $3$, $t_{\text{Log}}$ for all $\BS$ is increased by $1$.

 \begin{table}[htp]
	\centering
	\begin{tabular}{ | c | c | c | c | c |}
		
		\hline
		\centering
		BS ID & $\BS_{1}$ & $\BS_{2}$ &$\BS_{3}$ & $\BS_{4}$ \\ \hline
		SNR level & $\ell_{1}$ &  $\ell_{2}$ & $\ell_{2}$  & $\ell_{1}$ \\ \hline
		$t_{\text{Log}}$ & 0 & 1 & 2 & 0\\ 
		\hline
		
	\end{tabular}
\label{tab1}
\end{table}   
\end{example}

 The SNR log table will be updated in all the CIs and all UE's locations. During the $\DM_{1}$ phase, the backup BS is selected based on the optimal policy of the RL algorithm, described in the next section. It worths to mention that if an appropriate backup BS is selected, the previous records in log table will not be checked. Therefore, if the agent is trained well enough, the probability that a proper backup BS is selected is high.

\section{Learning Framework}
\label{learning}
The performance of the proposed handover approach heavily depends on how to select a backup BS in various CIs. This selection depends on the predictions of the SNR values and blockage of the BSs in future CIs. Such predictions, however, require a very detailed modeling of formidable complexity due to dynamicity of the obstacles and the UE mobility in mmWave networks. To address this problem, we use the RL framework to optimize the list of the backup BSs in various CIs.

The RL problem consists of  a set of environment states $\mathcal{S}$, a set of actions $\mathcal{A}(s)$, a set of rewards $\mathcal{R}\subset\mathbb{R}$, and transition probabilities that determine the next state based on the current state and action  \cite{sutton1998introduction}. In our problem, the transition probabilities model the SNR variations due to the UE's mobility through its trajectory and obstacle topology.
The agent is the decision maker (which can sit in the edge cloud) based on the policy. More details regarding RL components are reported in Appendix~B.

As it is shown in Fig. \ref{fig:fig1}, all the $\BS$s in a zone are connected to the agent. We define states as tuple $s=(s^{(1)}, s^{(2)}, s^{(3)})\in \mathcal{S}$ where $s^{(1)}\in [M]$ is the current location of the UE through the trajectory with length $M$, $s^{(2)}\in [N]$ is the index of the serving $\BS$s in the zone and $s^{(3)}\in [L]$ is the quantized SNR levels. The agent's action, $a \in \mathcal{A} = [N]$, is choosing a $\BS$ in order to ping as the backup $\BS$. The SNR of the backup channel is estimated in mini-time $B$. 

We define the agent's instantaneous reward as UE's achieved rate at the end of the CI as
\begin{equation*}
r(s_{i})= \R(i)
\end{equation*}
 and agent's long-term reward is the UE's
 trajectory rate ($\R_{\text{traj}}$) as
\begin{equation}
\R_{\text{traj}}=\sum_{i=1}^M \R(i).
\end{equation}

\begin{algorithm}[tp] 
	\caption{Handover. }\label{Alg1}
	\textbf{Inputs:} UE's mobility model including current location of the trajectory $i\in[M]$, number of $\BS$s in the UE's zone ($N$) and handover threshold ($\T_{\HO}$).
	\begin{algorithmic}[1]
		\State{Initialization: For $k=1$ set $\BS^{1}_{S}=\BS_{1}$}
		
		\State{ $t_{\text{Log}}(\BS_{j})=+\infty$ , \:  for all $j\in[N]$}
		
		\For {$i=1,...,M$}
		\State {$k$ $\mapsfrom$ current CI}
		\State 
		{// During mini-time slot $S$}
		
		\State 
		{Estimate channel from $\BS^{k}_{S}$ toward location $i$ and} 
		\State 
		{calculate the $\SNR(\BS^{k}_{S},i)$ level}  \State {Set $t_{\text{Log}}(\BS^{k}_{S})=0$}
		\State
		{// During mini-time slot $B$}
		
		\State {Choose $\BS^{k}_{B}=\BS_{j}$, $ j \in [N] $ based on $\DM_{1}$}
		\State
		{Estimate channel from $\BS^{k}_{B}$  toward location $i$ and }
		\State
		{	calculate the $\SNR(\BS^{k}_{B} ,i)$ level}
		\State
		{ Set $t_{\text{Log}}(\BS^{k}_{B} )=0$}
		\If{ $\SNR^\ell(\BS^{k}_{S},i)$ $> \T_{\HO} $ }
		\State
		{$\BS^{k+1}_{S}=\BS^k_{S}$} 	 
		
		\Else
		
		\State { // Perform handover}
		\If{ $\SNR^{\ell}(\BS^{k}_{B},i) > \T_{\HO} $ }
		\State {$\BS^{k+1}_{S}=\BS^k_{B}$} 	
		
		\Else
		
		\State {$\BS^{k+1}_{S}=\BS_{\acute{j}}$, where}  \State {$\acute{j}=\mathrm{argmin}_{j} \: t_{\text{Log}}(\BS_{j})$  s.t. $\SNR^{\ell}(j ,i)>\T_{\HO}$ }

		\EndIf
		\EndIf
		\State{	$t_{\text{Log}}(\BS_{j})=t_{\text{Log}}(\BS_{j})+1$, \:  for all $j\in[N]$} 
		
		\EndFor
		
		\\
		\textbf{Outputs:} $\BS_{S}$ and $\R_{i}$ \:  for all $i\in[M]$  
	\end{algorithmic}
\end{algorithm}
Therefore, the aim of the RL algorithm is to find the optimal policy ($\pi^\ast$) that maximizes the total UE's achieved rate through its trajectory, i.e.,
\begin{equation}
\pi^\ast=\argmax_{\pi}\mathbf{E} \: [\R_{\text{traj}}],
\label{policy}
\end{equation}
where the expectation is with respect to the randomness in the channel gain (fading and blockage).

In our case $\pi$ is a function from $\mathcal{S} \subseteq R$ to $\mathcal{A} = [N]$, i.e.,
 $$\pi: [M] \times [N] \times [L] \mapsto [N].$$
 So, $\pi$ is a 3-dimensional array which shows the best choice of backup $\BS$ for each state. 

 In order to find $\pi^\ast$, we use Q-learning algorithm, which 
 enables learning with no prior knowledge of the environment and finds the optimal decision based on the interactions with the environment, using Algorithm \ref{Alg2} \cite{sutton1998introduction}. In Appendix~C, we have provided more detailed information on the Q-learning and how it works.



		
		

\begin{algorithm}[t] 
	\caption{Q-Learning in $\epsilon$-greedy policy in one episode \cite{sutton1998introduction}}\label{Alg2}
	\begin{algorithmic}[1]
		\State{Initialization: An initial value $Q(s,a)  ~\forall s\in \mathcal{S} =[M] \times [N] \times [L], \forall a\in \mathcal{A}(s)=[N] $}
		
		\For {$i \in[M]$ }
		\State {Observe $s_{i}$ }
		\State{Take a random variable $\tau$ uniformly from [0,1]}
		\If{$\tau\leq \epsilon$ }
		\State{ Take a random action $a_{i}$ uniformly  from set $\mathcal{A}(s)$ }
		
		\Else
		\State{Take action $a_{i}=\argmax_{a \in \mathcal{A}(s)} Q^\ast(s,a)$  } 
		\EndIf	
		\State{Observe $s_{i+1}$ and $r(s_{i})$}
		\State {Update the action-value function as: } 
		
		\State { $Q(s_{i},a_{i})\leftarrow  Q(s_{i},a_{i})+\alpha [r(s_{i})+\gamma \max _{a \in \mathcal{A}(s_{i+1})}Q(s_{i+1},a)-Q(s_{i},a_{i})]$}

		\EndFor
		
	\end{algorithmic}
\end{algorithm}
\section{Simulation Results}
\label{simulation}
\begin{figure}[t]
	\centering
	\includegraphics[width=0.35\textwidth]{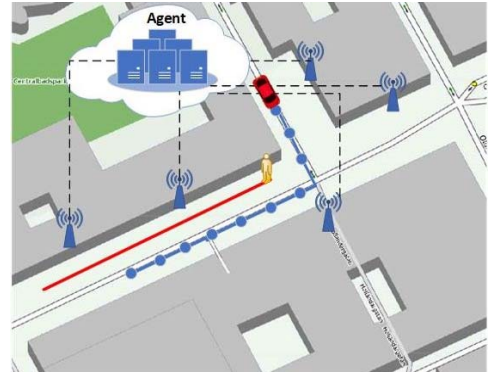}
	\caption{Simulation area. The red line shows the trajectory of a pedestrian. The blue line shows a vehicular trajectory whose location indexes are denoted by circles. The dash lines illustrate that all of the BSs are connected to the agent.}
	\label{fig:fig1}
\end{figure}

In this section, we present the performance evaluation of our proposed method in compared  to state-of-the-art benchmarks. 
We consider the downlink of mmWave network operating at $28$ GHz in two parts. In the first part, we use the narrow band cluster 3D channel model with different $\BS$ densities and in the second part, we use ray tracing tool to simulate a more realistic blockage and mmWave channel model. In all simulations, we fix the UE's trajectory. We consider a zone of $100 \times 100 ~ m^2$ area. The topology and the trajectories are shown in Fig. \ref{fig:fig1}. The simulation parameters are listed in Table \ref{table2}.

We compare the performance of our proposed handover method with  two baselines: \textit{multi-connectivity} handover \cite{dual2} and \textit{SMART} handover \cite{learninhRL}. In the \textit{multi-connectivity} handover baseline, the UE constantly checks the SNR of all the BS links in the zone and when experiences the blockage of the serving BS link, selects a BS with the highest quality. Although this baseline may provide an upper bound for the quality of the service \cite{dual2}, it suffers from the high computational complexity. The
\textit{SMART} baseline is based on an RL handoff policy which reduces the number of the handovers while keeps the UE's quality-of-service. In this baseline,  the BS selection algorithm is based on Upper Confidence Bound (UCB) algorithm \cite{auer2002finite} which has low complexity and achieves the optimal solution asymptotically. The UCB algorithm estimates  $\mathbf{E}[r(s,a)]$ by uniformly averaging the previously received rewards in state $s$ and action $a$. Then, it solves a set of equations which their solutions asymptotically converge to the solution of the Bellman equation. In \textit{SMART} algorithm the reward is the rate of the UEs and the set of equations are obtained by adding a term to the reward \cite{learninhRL}.

To evaluate the performance of our proposed beamforming method, we use the approach of \cite{8057188} as a baseline. In this baseline, the path skeleton sets are updated based on a fixed Euclidean distance (ED).
However, in our beamforming method the optimal path skeleton distance threshold ($\T_{\D}$) is found based on Equation (\refeq{T}). 


%

\begin{table}[t]
	\begin{center}
		\caption{Simulation parameters. }
		\label{table2}
		\begin{tabular}{|c|c|} 
			\hline
			\textbf{Parameters} & \textbf{Values in Simulations} \\
			\hline
			\hline 
			$\BS$ transmit power & $30$ dBm\\
			$\sigma^2$	&$-174$ dBm/Hz\\
			$W$ & $500$ MHz \\
			$	N_{\BS}$&$8\times 8$\\
			$ N_{\UE}$& $ 4\times 4$\\
			$\hat{n}_\LoS$	 & $3$ \\
			$\hat{n}_\NLoS$&$4$ \\
			\hline
			\multicolumn{2}{|c|}{Ray tracing parameters
			} \\
			\hline
			BS height	 & 6 m\\
			Brick penetration loss \cite{penetration}& 28.3 dB\\
			Glass penetration loss \cite{penetration}& 3.9 dB\\
			\hline
			\multicolumn{2}{|c|}{learning parameters
			} \\
			\hline
			$\alpha$&	$ 0.1$ \\
			$\gamma$&	$ 0.99$   \\
			$\epsilon$&	$0.01$ \\
			\hline 
		\end{tabular}
	\end{center}
\end{table}

\subsection{Narrow Band Cluster Channel Model}

In this model, we generate the mmWave links as described in Section \ref{model}. We consider a UE with a trajectory of $100$ m and two mobility models in our numerical studies: pedestrian and vehicle, where we assumed a UE speed of $5$ km/h and $36$ km/h, respectively. Due to the similarity of the results, we only report the pedestrian mobility model in the following.

We use $10^5$ different channel realizations as the input of the RL algorithm. During the learning phase, we run our algorithm  in order to reach the optimal policy that in this case, is the selecting optimal backup $\BS$ in different states\footnote{In our case, in order to converge to the optimal solution, we run our algorithm $10^6$ times. The running time by using a standard desktop computer is around $6$ hours.}. The handover threshold is $\T_{\HO}= 40$ dB. Based on the speed of the pedestrian, we define location indexes  in every $2$ m (50 location indexes  through the trajectory).
We fix the maximum number of path skeleton updated ($U_{\max}$) to $10$. We choose the distance between two consecutive updates in the ED approach equals to $10$ m to keep the same $U_{\max}$ as our approach.
We study two different scenarios with different $\BS$ densities as follows:


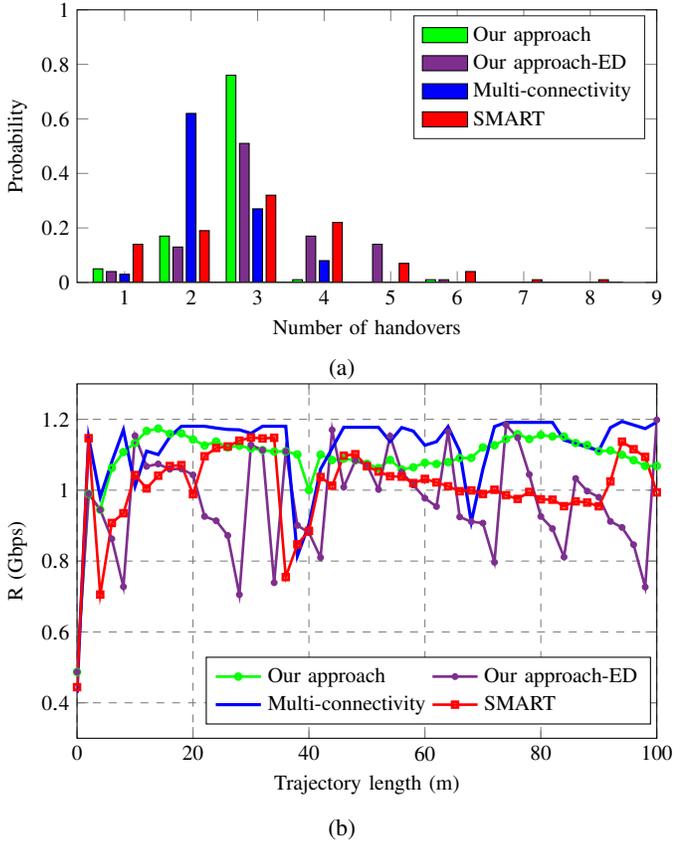
\begin{figure}[t]
	\centering
	\begin{subfigure}[t]{0.5\textwidth}
			{\footnotesize 
%
%
\definecolor{mycolor1}{rgb}{0.00000,0.44700,0.74100}%
\definecolor{mycolor2}{rgb}{0.85000,0.32500,0.09800}%
\definecolor{mycolor3}{rgb}{0.92900,0.69400,0.12500}%
\definecolor{mycolor4}{rgb}{0.49400,0.18400,0.55600}%
\begin{tikzpicture}

\begin{axis}[%
width=0.85\columnwidth,
height=0.4\columnwidth,
at={(0\columnwidth,0\columnwidth)},
scale only axis,
xmin=0.284999999999999,
xmax=9,
xlabel={Number of handovers},
ymin=0,
ymax=1,
ylabel style={},
ylabel={Probability},
axis background/.style={fill=white},
legend style={legend cell align=left, align=left, draw=white!15!black}
]
\addplot[ybar, bar width=0.145, fill=green, draw=black, area legend] table[row sep=crcr] {%
0.6	0.05\\
1.6	0.17\\
2.6 0.76\\
3.6	0.01\\
4.6	0\\
5.6	0.01\\
6.6	0\\
7.6	0\\
};
\addplot[forget plot, color=white!15!black] table[row sep=crcr] {%
0.509090909090909	0\\
8.49090909090909	0\\
};
\addlegendentry{Our approach}

\addplot[ybar, bar width=0.145, fill=mycolor4, draw=black, area legend] table[row sep=crcr] {%
0.8	0.04\\
1.8	0.13\\
2.8	0.51\\
3.8	0.17\\
4.8	0.14\\
5.8	0.01\\
6.8	0\\
7.8	0\\
};
\addplot[forget plot, color=white!15!black] table[row sep=crcr] {%
0.509090909090909	0\\
8.49090909090909	0\\
};
\addlegendentry{Our approach-ED}

\addplot[ybar, bar width=0.145, fill=blue, draw=black, area legend] table[row sep=crcr] {%
1	0.03\\
2	0.62\\
3	0.27\\
4	0.08\\
5	0\\
6	0\\
7	0\\
8	0\\
};
\addplot[forget plot, color=white!15!black] table[row sep=crcr] {%
0.509090909090909	0\\
8.49090909090909	0\\
};
\addlegendentry{Multi-connectivity}

\addplot[ybar, bar width=0.145, fill=red, draw=black, area legend] table[row sep=crcr] {%
1.2	0.14\\
2.2	0.19\\
3.2	0.32\\
4.2	0.22\\
5.2	0.07\\
6.2	0.04\\
7.2	0.01\\
8.2	0.01\\
};
\addplot[forget plot, color=white!15!black] table[row sep=crcr] {%
0.509090909090909	0\\
8.49090909090909	0\\
};
\addlegendentry{SMART}

\end{axis}
\end{tikzpicture}%
}
		\caption{}
		\label{fig6:(a)}
	\end{subfigure}
	\begin{subfigure}[t]{0.5\textwidth}
		{\footnotesize 
%
%
\definecolor{mycolor1}{rgb}{0.00000,0.44700,0.74100}%
\definecolor{mycolor2}{rgb}{0.85000,0.32500,0.09800}%
\definecolor{mycolor3}{rgb}{0.92900,0.69400,0.12500}%
\definecolor{mycolor4}{rgb}{0.49400,0.18400,0.55600}%
\pgfplotsset{major grid style={dashed, gray}}
\begin{tikzpicture}

\begin{axis}[%
width=0.85\columnwidth,
height=0.52\columnwidth,
at={(0\columnwidth,0\columnwidth)},
scale only axis,
xmin=0,
xmax=100,
xlabel style={},
xlabel={Trajectory length (m)},
ymin=0.3,
ymax=1.3,
ylabel style={},
ylabel={R (Gbps)},
axis background/.style={fill=white},
xmajorgrids,
ymajorgrids,
legend style={at={(0.99,0.23)},anchor=north east,legend cell align=left,fill =white,align=right,draw=black,legend columns=2}
]
\addplot [color=green, line width=1.0pt, mark=o, mark options={mark size=1.1pt, solid, green}]
  table[row sep=crcr]{%
0	0.48700151381658\\
2	0.989846988347938\\
4	0.944714288824133\\
6	1.06267138980105\\
8	1.10761058257832\\
10	1.13339836341886\\
12	1.16716874338495\\
14	1.17395627319864\\
16	1.15913820371729\\
18	1.1601819151557\\
20	1.14307437034072\\
22	1.12599699581329\\
24	1.13735488309401\\
26	1.12045942808645\\
28	1.12536575514847\\
30	1.11800379036344\\
32	1.11307538256344\\
34	1.10892379457025\\
36	1.1094233008037\\
38	1.10071805915563\\
40	1.00006666959721\\
42	1.10035935484029\\
44	1.0851120775027\\
46	1.08871331712078\\
48	1.08435209614234\\
50	1.07454801485074\\
52	1.06222737763366\\
54	1.085267491325\\
56	1.05842532765156\\
58	1.06452719326246\\
60	1.07734345991792\\
62	1.07383791615045\\
64	1.07911388989639\\
66	1.09045444334075\\
68	1.09111981181071\\
70	1.1207134859332\\
72	1.12665219940671\\
74	1.14421888395011\\
76	1.15796463307543\\
78	1.14433136639886\\
80	1.15573778516688\\
82	1.15150859987436\\
84	1.15142946470763\\
86	1.133013068475\\
88	1.12734582370844\\
90	1.11004940376204\\
92	1.11183186630324\\
94	1.09987691458485\\
96	1.08461566659803\\
98	1.06826768757507\\
100	1.06828507798124\\
};
\addlegendentry{Our approach}

\addplot
[color=mycolor4, line width=1.0pt, mark=*, mark options={mark size=0.8pt, solid, =mycolor4}]
  table[row sep=crcr]{%
0	0.48700151381658\\
2	0.989846988347938\\
4	0.944714288824133\\
6	0.86267138980105\\
8	0.72761058257832\\
10	1.15339836341886\\
12	1.06716874338495\\
14	1.07395627319864\\
16	1.05913820371729\\
18	1.0601819151557\\
20	1.04307437034072\\
22	0.92599699581329\\
24	0.913735488309401\\
26	0.872045942808645\\
28	0.70536575514847\\
30	1.12800379036344\\
32	1.11307538256344\\
34	0.73892379457025\\
36	1.1094233008037\\
38	0.90071805915563\\
40	0.880006666959721\\
42	0.810035935484029\\
44	1.1701120775027\\
46	1.00871331712078\\
48	1.08335209614234\\
50	1.07154801485074\\
52	1.00222737763366\\
54	1.15267491325\\
56	1.05042532765156\\
58	1.01452719326246\\
60	0.97734345991792\\
62	0.95383791615045\\
64	1.165000388989639\\
66	0.9245444334075\\
68	0.9111981181071\\
70	0.907134859332\\
72	0.79665219940671\\
74	1.18421888395011\\
76	1.14796463307543\\
78	1.04433136639886\\
80	0.92573778516688\\
82	0.89150859987436\\
84	0.81142946470763\\
86	1.033013068475\\
88	0.99734582370844\\
90	0.98004940376204\\
92	0.91183186630324\\
94	0.89487691458485\\
96	0.8461566659803\\
98	0.726768757507\\
100	1.19828507798124\\
};
\addlegendentry{Our approach-ED}

\addplot [color=blue, line width=1.2pt]
  table[row sep=crcr]{%
0	0.427326958889994\\
2	1.14820051702003\\
4	0.980328569829208\\
6	1.08032856982921\\
8	1.17032856982921\\
10	1.01032856982921\\
12	1.11032856982921\\
14	1.10032856982921\\
16	1.15032856982921\\
18	1.18032856982921\\
20	1.18032856982921\\
22	1.18032856982921\\
24	1.17532856982921\\
26	1.17132856982921\\
28	1.17032856982921\\
30	1.16032856982921\\
32	1.18032856982921\\
34	1.18032856982921\\
36	1.18032856982921\\
38	0.81407281005836\\
40	0.90407281005836\\
42	1.05940728100584\\
44	1.11831107623719\\
46	1.17747957001881\\
48	1.17747957001881\\
50	1.17747957001881\\
52	1.17731203427522\\
54	1.13731203427522\\
56	1.17671567028442\\
58	1.16671567028442\\
60	1.12671567028442\\
62	1.13671567028442\\
64	1.17915310779737\\
66	1.11053609972421\\
68	0.910300985436351\\
70	1.0593872125717\\
72	1.1793872125717\\
74	1.1913373277047\\
76	1.1913373277047\\
78	1.1913373277047\\
80	1.1913373277047\\
82	1.1913373277047\\
84	1.1413373277047\\
86	1.1313373277047\\
88	1.1213373277047\\
90	1.1113373277047\\
92	1.17649065260115\\
94	1.19395057467383\\
96	1.18395057467383\\
98	1.17367212378342\\
100	1.19209064437003\\
};
\addlegendentry{Multi-connectivity}

\addplot [color=red, line width=1.0pt, mark=square, mark options={mark size=1.1pt, solid, red}]
  table[row sep=crcr]{%
0	0.44408376346766\\
2	1.14669573921479\\
4	0.70541527544144\\
6	0.907429824553192\\
8	0.934971918780911\\
10	1.04277401629135\\
12	1.00492027311335\\
14	1.04082440206857\\
16	1.06805960625608\\
18	1.07073155178501\\
20	0.9886889261973\\
22	1.09562392120693\\
24	1.11915442213663\\
26	1.12336286235677\\
28	1.13975820335618\\
30	1.14858243494468\\
32	1.14591606710326\\
34	1.14790396171745\\
36	0.754799813385964\\
38	0.847145294238311\\
40	0.8849996516008\\
42	1.03696705323705\\
44	1.012747570072\\
46	1.09741930775588\\
48	1.10150587494002\\
50	1.06736819280085\\
52	1.05252148355127\\
54	1.03925486277634\\
56	1.03803936713085\\
58	1.02004198868424\\
60	1.03178942558296\\
62	1.02185649292866\\
64	1.01098273766464\\
66	0.99705841429989\\
68	0.999280883241391\\
70	0.989220401788804\\
72	1.00115917157652\\
74	0.985732722729782\\
76	0.975223122518869\\
78	0.995001835984698\\
80	0.974224231241359\\
82	0.973177946542439\\
84	0.954873140737573\\
86	0.968399431967442\\
88	0.965055455932907\\
90	0.954627598507202\\
92	1.02466131083527\\
94	1.13693245181147\\
96	1.11561048905346\\
98	1.09393407846518\\
100	0.99377443751082\\
};
\addlegendentry{SMART}

\end{axis}
\end{tikzpicture}%
}
		\caption{}
		\label{fig6:(b)}
	\end{subfigure}
	
		
	\caption{Handover performance of our approach compared to baselines in a sparse mmWave network.}
	\label{fig6:6}
\end{figure}
\subsubsection{First Scenario (sparse mmWave network)}
In the first scenario, we consider a $\BS$ density of $\rho=5\times10^{-4}/m^{2}$, which corresponds to the average of one $\BS$ in every $50 ~ m^2$.
 After finding the optimal policy, we run our proposed algorithm over $10^4$ different channel realizations. Fig. \ref{fig6:6} compares the performance of our approach to the baselines.
 In \textit{our approach}, we use our proposed beamforming method and handover algorithm. In \textit{our approach-ED}, we use the ED approach to trigger beamforming in our handover algorithm. In \textit{multi-connectivity} and \textit{SMART} baselines, we apply our skeleton-based beamforming method while using their handover algorithms.
  In particular, Fig. \ref{fig6:(a)} shows the number of handovers when the UE moves along the trajectory, and \mbox{Fig. \ref{fig6:(b)}} shows the average $\R(i)$, $i\in[100]$. The average trajectory rate, $\R_{\text{traj}}$, in this scenario is shown in Table III. It can be seen from the figure that our proposed learning-based handover  provides almost a same rate in all locations of the trajectory with a small number of the handovers. Even in the case of handover, our approach can maintain an almost constant rate for the UE, making it suitable for services with high reliability requirements. Moreover, based on Table III, our approach provides high trajectory rate  by considering the long term effects of the handovers. The baselines, however, suffer from either high number of handovers or high fluctuations on $\R_{i}$.
  In \textit{our approach-ED}, non-optimal  path skeleton updating causes higher rate fluctuations and higher number of handover needed  compared to \textit{our approach}.
  In the case of \textit{Multi-connectivity} baseline, the additional connection-probe procedure increases the complexity of the method from the UE's perspective.

\subsubsection{Second Scenario (dense mmWave network)}

\begin{figure}[t]
	\centering
	\begin{subfigure}[t]{0.5\textwidth}
			{\footnotesize 
%
%
\definecolor{mycolor1}{rgb}{0.00000,0.44700,0.74100}%
\definecolor{mycolor2}{rgb}{0.85000,0.32500,0.09800}%
\definecolor{mycolor3}{rgb}{0.92900,0.69400,0.12500}%
\definecolor{mycolor4}{rgb}{0.49400,0.18400,0.55600}%
\begin{tikzpicture}

\begin{axis}[%
width=0.85\columnwidth,
height=0.4\columnwidth,
at={(0\columnwidth,0\columnwidth)},
scale only axis,
xmin=0,
xmax=9,
xlabel style={},
xlabel={Number of handovers},
ymin=0,
ymax=1,
ylabel style={},
ylabel={Probability},
axis background/.style={fill=white},
legend style={at={(0.99,0.99)},anchor=north east,legend cell align=left,fill =white,align=right,draw=black,legend columns=1}
]
\addplot[ybar, bar width=0.145, fill=green, draw=black, area legend] table[row sep=crcr] {%
0.6	0.72\\
1.6	0.12\\
2.6	0.15\\
3.6	0.01\\
4.6	0\\
5.6	0\\
6.6	0\\
};
\addplot[forget plot, color=white!15!black] table[row sep=crcr] {%
0.509090909090909	0\\
7.49090909090909	0\\
};
\addlegendentry{Our approach}

\addplot[ybar, bar width=0.145, fill=mycolor4, draw=black, area legend] table[row sep=crcr] {%
0.8	0.12\\
1.8	0.1\\
2.8	0.5 \\
3.8	0.28\\
4.8	0\\
5.8	0\\
6.8	0\\
};
\addplot[forget plot, color=white!15!black] table[row sep=crcr] {%
0.509090909090909	0\\
7.49090909090909	0\\
};
\addlegendentry{Our approach-ED}

\addplot[ybar, bar width=0.145, fill=blue, draw=black, area legend] table[row sep=crcr] {%
1	0.49\\
2	0.48\\
3	0.03\\
4	0\\
5	0\\
6	0\\
7	0\\
};
\addplot[forget plot, color=white!15!black] table[row sep=crcr] {%
0.509090909090909	0\\
7.49090909090909	0\\
};
\addlegendentry{Multi-connectivity}

\addplot[ybar, bar width=0.145, fill=red, draw=black, area legend] table[row sep=crcr] {%
1.2	0.04\\
2.2	0.35\\
3.3	0.32\\
4.2	0.17\\
5.2	0.08\\
6.2	0.03\\
7.2	0.01\\
};
\addplot[forget plot, color=white!15!black] table[row sep=crcr] {%
0.509090909090909	0\\
7.49090909090909	0\\
};
\addlegendentry{SMART}

\end{axis}
\end{tikzpicture}%
}
		\caption{}
		\label{fig5:(a)}
	\end{subfigure}
	\begin{subfigure}[t]{0.5\textwidth}
			{\footnotesize 
%
%
\definecolor{mycolor1}{rgb}{0.00000,0.44700,0.74100}%
\definecolor{mycolor2}{rgb}{0.85000,0.32500,0.09800}%
\definecolor{mycolor3}{rgb}{0.92900,0.69400,0.12500}%
\definecolor{mycolor4}{rgb}{0.49400,0.21400,0.55600}%
\pgfplotsset{major grid style={dashed, gray}}
\begin{tikzpicture}

\begin{axis}[%
width=0.85\columnwidth,
height=0.52\columnwidth,
at={(0\columnwidth,0\columnwidth)},
scale only axis,
xmin=0,
xmax=100,
xlabel style={},
xlabel={Trajectory length (m)},
ymin=0.5,
ymax=1.15,
ylabel style={},
ylabel={R (Gbps)},
axis background/.style={fill=white},
xmajorgrids,
ymajorgrids,
legend style={at={(0.99,0.23)},anchor=north east,legend cell align=left,fill =white,align=right,draw=black,legend columns=2}
]
\addplot [color=green, line width=1pt, mark=o, mark options={mark size=1.1pt,solid, green}]
table[row sep=crcr]{%
	0	1.04012430157605\\
	2	1.05695552474032\\
	4	1.06616718694681\\
	6	1.05737673377037\\
	8	1.06573606041023\\
	10	1.05101903413072\\
	12	1.04459644248302\\
	14	1.04658968646241\\
	16	1.03464573894146\\
	18	1.04194005102057\\
	20	1.03908227229881\\
	22	1.02578519155179\\
	24	1.02882916362866\\
	26	1.01047194910707\\
	28	0.995483633994885\\
	30	1.02220860231231\\
	32	1.00205974775317\\
	34	0.995457984955467\\
	36	1.01140859863378\\
	38	1.00867005676891\\
	40	0.994928669506726\\
	42	1.02009757999883\\
	44	1.01055041461194\\
	46	1.01714689182129\\
	48	0.995086823342193\\
	50	1.00460999352576\\
	52	1.01160560991865\\
	54	0.991180444994401\\
	56	1.00103671284949\\
	58	0.990992907872306\\
	60	0.965880916954906\\
	62	0.992517165598616\\
	64	1.01577184227245\\
	66	1.00628866717702\\
	68	0.995231865873943\\
	70	1.00585657800548\\
	72	1.0240641113377\\
	74	0.978651383913395\\
	76	1.00837162180786\\
	78	1.00650211142862\\
	80	1.00440813269226\\
	82	1.00046781825263\\
	84	1.01435637002435\\
	86	1.03218950922673\\
	88	1.03131928753875\\
	90	1.0328370197477\\
	92	1.04477387813015\\
	94	1.02042735938067\\
	96	1.03916951727058\\
	98	1.00060400542939\\
	100	1.0076897844269\\
};
\addlegendentry{Our approach}

\addplot 
 [color=mycolor4, line width=1.0pt, mark=*, mark options={mark size=0.8pt, solid, =mycolor4}]
  table[row sep=crcr]{%
0	1.04012430157605\\
2	0.95695552474032\\
4	0.8616718694681\\
6	0.75737673377037\\
8	0.69573606041023\\
10	1.05801903413072\\
12	1.03459644248302\\
14	1.03158968646241\\
16	1.02464573894146\\
18	1.051194005102057\\
20	0.993908227229881\\
22	0.842578519155179\\
24	0.72882916362866\\
26	0.7101047194910707\\
28	1.045483633994885\\
30	1.00220860231231\\
32	0.90205974775317\\
34	0.895457984955467\\
36	0.86140859863378\\
38	0.850867005676891\\
40	0.9804928669506726\\
42	0.92009757999883\\
44	0.901055041461194\\
46	0.90014689182129\\
48	0.895086823342193\\
50	1.00460999352576\\
52	1.0000560991865\\
54	0.801180444994401\\
56	1.0000103671284949\\
58	0.890992907872306\\
60	0.815880916954906\\
62	0.792517165598616\\
64	0.751577184227245\\
66	1.01628866717702\\
68	0.975231865873943\\
70	0.90585657800548\\
72	0.89240641113377\\
74	0.8578651383913395\\
76	1.00837162180786\\
78	1.00050211142862\\
80	1.01440813269226\\
82	1.02046781825263\\
84	1.01435637002435\\
86	0.99218950922673\\
88	1.0003131928753875\\
90	1.0008370197477\\
92	0.8704477387813015\\
94	0.822042735938067\\
96	0.793916951727058\\
98	0.80060400542939\\
100	0.89076897844269\\
};
\addlegendentry{Our approach-ED}

\addplot [color=blue, line width=1.0pt]
table[row sep=crcr]{%
	0	1.04316351136866\\
	2	1.05431635113687\\
	4	1.11316351136866\\
	6	0.89453\\
	8	0.843163511368662\\
	10	0.813163511368662\\
	12	0.9220775615694\\
	14	0.8920775615694\\
	16	1.00220775615694\\
	18	1.04220775615694\\
	20	1.04135900322496\\
	22	1.04135900322496\\
	24	1.04098778264593\\
	26	1.02098778264593\\
	28	1.04098778264593\\
	30	1.06098778264593\\
	32	1.01098778264593\\
	34	1.02098778264593\\
	36	1.030782645925\\
	38	0.940987782645925\\
	40	0.990987782645925\\
	42	1.0415683636776\\
	44	1.0415683636776\\
	46	1.0315683636776\\
	48	1.0215683636776\\
	50	1.0415683636776\\
	52	0.913480296550813\\
	54	0.943480296550813\\
	56	1.00348029655081\\
	58	1.0180296550813\\
	60	1.09551347124114\\
	62	1.09582191132708\\
	64	1.09846451889401\\
	66	0.827505000876119\\
	68	0.97505000876119\\
	70	1.01750500087612\\
	72	0.880097382102374\\
	74	1.09726794247141\\
	76	1.09726794247141\\
	78	1.09726794247141\\
	80	1.09726794247141\\
	82	1.08899478061524\\
	84	1.09199478061524\\
	86	1.09983030924465\\
	88	1.10058142726591\\
	90	1.1012337381805\\
	92	1.10114857100481\\
	94	1.00383665973419\\
	96	1.10383665973419\\
	98	1.10917378351021\\
	100	1.10866303210156\\
};
\addlegendentry{Multi-connectivity}

\addplot [color=red, line width=1.0pt, mark size=1.3pt , mark=square, mark options={mark size=1.1pt,solid, red}]
table[row sep=crcr]{%
	0	1.04318341369049\\
	2	1.04814079397336\\
	4	1.05891221691462\\
	6	1.06498162726106\\
	8	1.04926159690158\\
	10	1.058617835386\\
	12	1.06297607415382\\
	14	1.05280000809919\\
	16	1.05786899438335\\
	18	1.05881257122623\\
	20	1.03448200734037\\
	22	1.02861331884921\\
	24	0.89693925956067\\
	26	0.83123456661346\\
	28	0.792675544398718\\
	30	0.9741107099315\\
	32	0.997720475734172\\
	34	0.995562203171324\\
	36	0.997663231901966\\
	38	0.978189512928692\\
	40	0.969212958798552\\
	42	0.987183519912955\\
	44	0.96004421777333\\
	46	0.968948437466146\\
	48	0.941899096308794\\
	50	0.954189771566717\\
	52	0.942472141145211\\
	54	0.952260870848154\\
	56	0.943895571968046\\
	58	0.941807674350539\\
	60	0.750603903514207\\
	62	0.964559854696534\\
	64	0.938789846904528\\
	66	0.976237890664157\\
	68	0.986088526953129\\
	70	0.982415892885393\\
	72	0.984340987164684\\
	74	0.919566861585762\\
	76	0.994300418082237\\
	78	0.995026538883377\\
	80	0.980546701530564\\
	82	0.977545485862282\\
	84	0.983729196074909\\
	86	0.990775543862681\\
	88	0.979273490180129\\
	90	0.968661070259166\\
	92	0.962954396784067\\
	94	0.890222272310907\\
	96	0.982283610466017\\
	98	0.961956227090807\\
	100	0.99169925001442\\
};
\addlegendentry{SMART}

\end{axis}
\end{tikzpicture}%
}
		
		\caption{}
		\label{fig5:(b)}
	\end{subfigure}
	
		
	\caption{Handover performance of our approach compared to baselines in a dense mmWave network. }
	\label{fig5:5}
\end{figure}
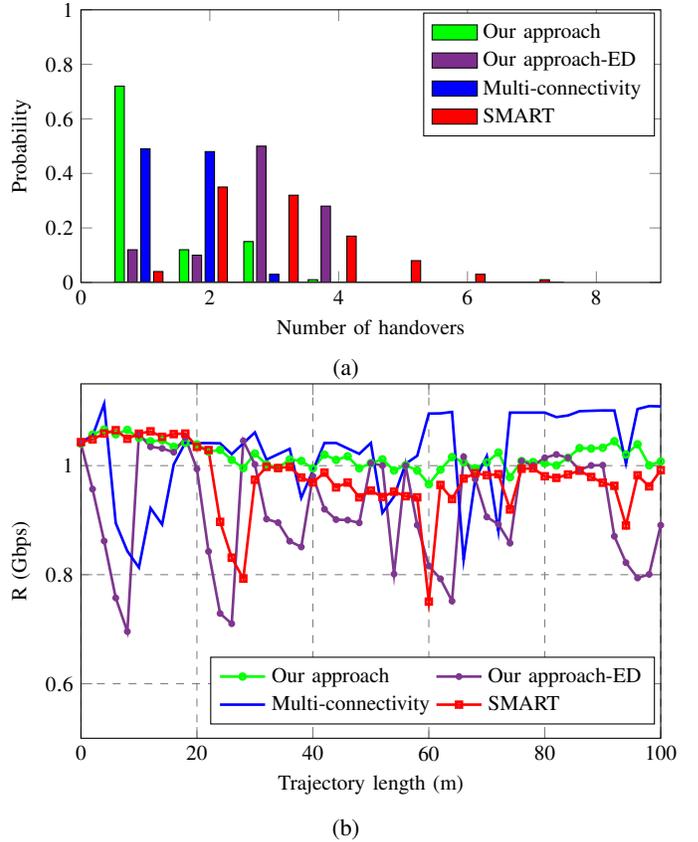

In the second scenario, the density of the $\BS$s is $\rho=10^{-3}/m^{2}$ which corresponds to the average of one $\BS$ in every $30 ~ m^2$. 
\mbox{Fig. \ref{fig5:(a)}} illustrates the number of the handovers, and \mbox{Fig. \ref{fig5:(b)}} shows the average $\R(i)$, $i\in[100]$.
It is evident that our learning-based handover method due to having less rate fluctuations along the trajectory is more reliable handover method specially when a high quality-of-service is needed. Furthermore, our method keeps the number of handover needed small  while providing comparable $\R(i)$ and $\R_{\text{traj}}$ (as it is shown in Table III) in comparison to other two baselines and \textit{our approach-ED}.

 \begin{table}[]
	\begin{center}
		\label{table3}
		\caption{Average trajectory rate, $\R_{\text{traj}}$, (Gbps).}
		\begin{tabular}{|c|c|c|} 
			\hline
			\textbf{Methods} & \textbf{Sparse scenario}  & \textbf{Dense scenario} \\
			\hline
			\hline
			Our approach & 55.65&51.90\\
			\hline
			Our approach-ED &	52.13 &	50.03\\
			\hline
			Multi-connectivity	&  57.08 & 52.17 \\
			\hline
			SMART	&  51.39 & 49.8062 \\
			\hline
		\end{tabular}
	\end{center}
\end{table}
\subsection{Ray Tracing}

We have also used ray tracing to model and evaluate the performance of our approach in a more realistic environment. In this part, we use our beamforming method as described in Section \ref{method}-A and evaluate the performance of our proposed handover algorithm. For the network topology, we extracted the real building map of the central part of Stockholm city from \textit{ open street map} data as the input of the ray tracing tool \cite{simic2017demo}. Then, we randomly assigned brick or glass materials to the buildings as the permanent obstacles. We also added some temporary random obstacles with height 1.5 m to model human bodies and some temporary random obstacles with width $4$ m and heights $1$ m and $3$ m to model various vehicles with different heights. The number, the position and the material loss of the temporary obstacles were chosen randomly in each realization of the channel. The simulation area is illustrated in Fig. \ref{fig:fig1}. UEs are moving along the real streets as extracted from  \textit{ open street map} data. We consider the UE's trajectory with length $100$ m and the vehicle mobility model. We consider the location indexes every $10$ m (11 location indexes through the trajectory) as it is shown with the blue line in Fig. \ref{fig:fig1}. The density of the $\BS$s is $\rho=5\times10^{-4}/m^{2}$. The handover threshold is $\T_{\HO}=0$ dB.

We run the ray tracing simulator for all the $\BS$s for different topologies. We considered each topology with different distribution and density of the temporary obstacles.

\begin{figure}[t]
	\centering
\begin{subfigure}[t]{0.5\textwidth}
	{\footnotesize 
%
%
\definecolor{mycolor1}{rgb}{0.00000,0.44700,0.74100}%
\definecolor{mycolor2}{rgb}{0.85000,0.32500,0.09800}%
\definecolor{mycolor3}{rgb}{0.92900,0.69400,0.12500}%
\begin{tikzpicture}

\begin{axis}[%
width=0.85\columnwidth,
height=0.4\columnwidth,
at={(0\columnwidth,0\columnwidth)},
scale only axis,
xmin=0,
xmax=8,
xlabel style={},
xlabel={Number of handovers},
ymin=0,
ymax=1,
ylabel style={},
ylabel={Probability},
axis background/.style={fill=white},
legend style={at={(0.99,0.99)},anchor=north east,legend cell align=left,fill =white,align=right,draw=black,legend columns=1}
]
\addplot[ybar, bar width=0.178, fill=green, draw=black, area legend] table[row sep=crcr] {%
0.8	0.666666666666667\\
1.8	0.333333333333333\\
3	0\\
4	0\\
5	0\\
6	0\\
};
\addplot[forget plot, color=white!15!black] table[row sep=crcr] {%
0.511111111111111	0\\
6.48888888888889	0\\
};
\addlegendentry{Our approach}

\addplot[ybar, bar width=0.178, fill=blue, draw=black, area legend] table[row sep=crcr] {%
1	0.533333333333333\\
2	0.466666666666667\\
3	0\\
4	0\\
5	0\\
6	0\\
};
\addplot[forget plot, color=white!15!black] table[row sep=crcr] {%
0.511111111111111	0\\
6.48888888888889	0\\
};
\addlegendentry{Multi-connectivity}

\addplot[ybar, bar width=0.178, fill=red, draw=black, area legend] table[row sep=crcr] {%
1.2	0.0333333333333333\\
2.2	0.333333333333333\\
3.2	0.166666666666667\\
4.2	0.266666666666667\\
5.2	0.133333333333333\\
6.2	0.0666666666666667\\
};
\addplot[forget plot, color=white!15!black] table[row sep=crcr] {%
0.511111111111111	0\\
6.48888888888889	0\\
};
\addlegendentry{SMART}

\end{axis}

\end{tikzpicture}%
}
	\caption{}
		\label{fig7:(a)}
\end{subfigure}
	\begin{subfigure}[t]{0.5\textwidth}
		{\footnotesize 
%
%
\pgfplotsset{major grid style={dashed, gray}}
\begin{tikzpicture}

\begin{axis}[%
width=0.85\columnwidth,
height=0.52\columnwidth,
at={(0\columnwidth,0\columnwidth)},
scale only axis,
xmin=1,
xmax=100,
xlabel style={},
xlabel={Trajectory length (m)},
ymin=0,
ymax=1.2,
ylabel style={},
ylabel={Rate (Gbps)},
axis background/.style={fill=white},
xmajorgrids,
ymajorgrids,
legend style={at={(0.99,0.32)},anchor=north east,legend cell align=left,fill =white,align=right,draw=black,legend columns=1}
]
\addplot [color=green, line width=1.0pt, mark=o, mark options={mark size=1.1pt,solid, green}]
  table[row sep=crcr]{%
1	0.75\\
10	0.8\\
20	0.75\\
30	0.9\\
40	0.86\\
50	0.79\\
60	0.71\\
70	0.8\\
80	0.89\\
90	0.71\\
100	0.78\\
};
\addlegendentry{Our aproach}

\addplot [color=blue, line width=1.0pt]
  table[row sep=crcr]{%
1	0.78\\
10	0.83\\
20	0.64\\
30	0.91\\
40	0.89\\
50	0.79\\
60	0.75\\
70	0.83\\
80	0.89\\
90	0.66\\
100	0.81\\
};
\addlegendentry{Multi-connectivity}

\addplot [color=red, line width=1.0pt , mark=square, mark options={mark size=1.1pt,solid, red}]
  table[row sep=crcr]{%
1	0.61\\
10	0.71\\
20	0.75\\
30	0.85\\
40	0.59\\
50	0.71\\
60	0.71\\
70	0.8\\
80	0.75\\
90	0.71\\
100	0.65\\
};
\addlegendentry{SMART}
\end{axis}
\end{tikzpicture}%
}
		\caption{}
		\label{fig7:(b)}
	\end{subfigure}
	
		
	\caption{Handover performance of our approach compared to baselines for simulated data. Topology is given by Fig. \ref{fig:fig1}.}
	\label{fig7:7}
\end{figure}
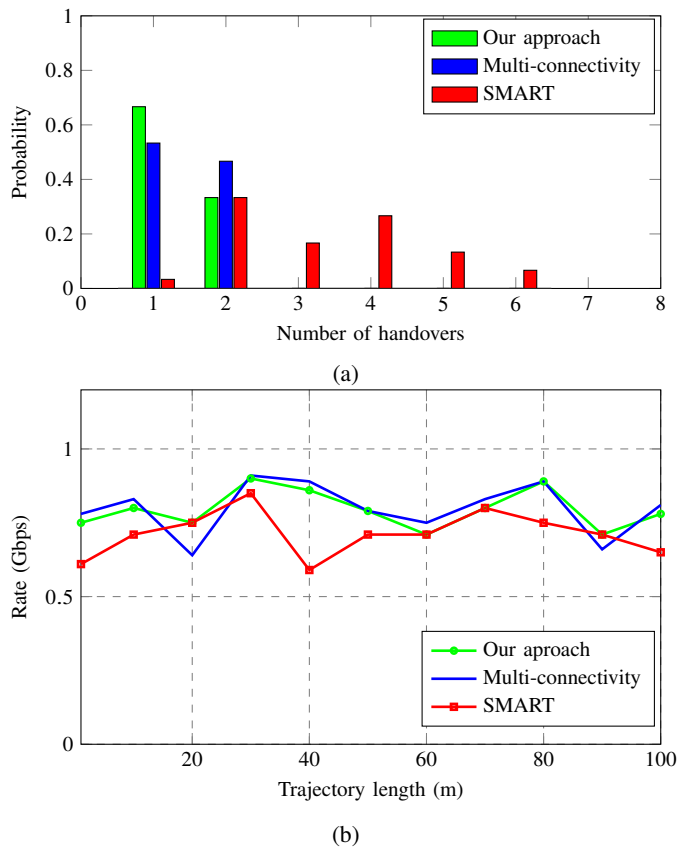

As it can be seen in Fig. \ref{fig7:7}, our proposed solution needs less number of the handovers while keeps the rate in each location almost consistent in comparison to other two baselines. Moreover, our approach by selecting the optimal actions in each state provides a near maximum long term reward ($\R_{\text{traj}}$) as it is shown in Table IV. 

\section{Conclusions}
 \label{conclusion} 

In this paper, we jointly considered the beamforming and handover challenges in mobile mmWave networks. We proposed an efficient beamforming method which leverages the sparsity and the spatial correlation of path skeletons in the channel estimation phase. 
We designed a handover algorithm based on a pilot structure that consists of an extra mini-slot regarding channel estimation toward backup BS. We used reinforcement learning algorithm as a decision maker regarding the choice of the backup BS in all locations of the UE's trajectory. Our proposed algorithm triggers a handover to a backup channel when  the link quality drops below a predefined threshold.
 We evaluated the performance of our method based on the narrow band cluster channel model and ray tracing. The results showed that our approach provides a reliable connection with the consistent rate through the UE's trajectory.

For future work, we plan to consider the load balancing and resource management in  our joint beamforming and handover approach. Moreover, we plan to work on an accurate mobility prediction scheme using the tracking and localization capabilities of the mmWave networks.

\appendices 
\section{}

	In order to solve \eqref{codebook}, we define matrices $\mathbf{G}\in\mathbb{C}^{\lvert W\rvert \times \lvert F\rvert}, \mathbf{W}\in\mathbb{C}^{d_w \times \lvert W\rvert}, \mathbf{F}\in\mathbb{C}^{d_f \times \lvert F \rvert}$ as following
	\begin{IEEEeqnarray*}{l}
		\mathbf{W} := [\mathbf{w}_1~~\mathbf{w}_2~~\ldots~~\mathbf{w}_{\lvert\mathcal{W}\rvert}], \\
		\mathbf{F} := [\mathbf{f}_1~~\mathbf{f}_2~~\ldots~~\mathbf{f}_{\lvert\mathcal{F}\rvert}], \\
		\mathbf{G} := \mathbf{W}^\HH \mathbf{H} \mathbf{F},
	\end{IEEEeqnarray*}
	where $d_w$ and $d_f$ are the dimension of codewords in $\mathcal{W}$ and $\mathcal{F}$, respectively; $\mathbf{w}_a\in\mathcal{W}$ and $\mathbf{f}_b\in\mathcal{F}$ are the codewords, so $\mathbf{W}$ and $\mathbf{F}$ are made by concatenating  all codewords.
	It can be obtained that for $\mathbf{w}_a\in\mathcal{W}$ and $\mathbf{f}_b\in\mathcal{F}$, $g_{a,b} = \mathbf{w}_a^\HH \mathbf{H} \mathbf{f}_b$.
	Hence, in order to find the solution of \eqref{codebook}, one only needs to find $a^*,b^*$ such that $\lvert g_{a^*,b^*}\rvert^2 \geq \lvert g_{a,b} \rvert^2$ for all $a\in [\lvert \mathcal{W} \rvert], b\in [\lvert \mathcal{F} \rvert]$.
	As a result, $\mathbf{w}_{a^*}, \mathbf{f}_{b^*}$ is the optimal solution of \eqref{codebook}.
  \begin{table}[t]
	\begin{center}
		\label{table4}
		\caption{Average trajectory rate, $\R_{\text{traj}}$, (Gbps).}
			\begin{tabular}{|c|c|} 
			\hline
			\textbf{Methods}  & \textbf{Ray tracing}\\
			\hline
			\hline
			Our approach& 8.74\\
			\hline
			Multi-connectivity &8.78\\
			\hline
			SMART&7.87\\
			\hline
		\end{tabular}
	\end{center}
\end{table}

\section{Reinforcement Learning} 
In this part, we introduce the reinforcement learning (RL) model as a decision making tool. 
At each step $k$, the agent makes a decision regarding a new action based on the immediate received reward of its action in the previous step.
The agent interacts with the environment in order to maximizes the reward while exploring the environment.
The main goal of the agent in RL is learning the optimal policy $\pi^\ast$ in order to maximize its long-term reward.
In particular, the agent maximizes the expected discounted reward and finds the optimal policy as \cite{sutton1998introduction}:
\begin{equation}
\pi^\ast = \argmax_{\pi} \mathbf{E}\left [\sum_{k=0}^\infty \gamma^k r(S_k, A_k) \Big| S_{0}=s \right ], 
\end{equation}
where $r(S_k, A_k)$ is the reward that the agent receives in time $k$ and $\gamma \in [0,1]$ is the discount factor.
The expectation is with respect to the randomness of the states. 

In episodic learning, the interactions between the agent and the environment is divided into subsequences of consecutive steps which are called episodes  \cite{sutton1998introduction}.
Each episode has a limited number of steps. In our scenario, each running of a fixed trajectory with a random distribution of the temporary obstacles is considered as an episode with length $M$ where $M$ is the length of the trajectory.

\section{Q-learning}
The optimal action-value function $Q^\ast(s,a)$ is defined as:
\begin{IEEEeqnarray}{rCl}
	Q^\ast(s,a)
	&=& \max_{\pi} \mathbf{E}\left [\sum_{k=1}^\infty \gamma^{k} r(S_k,A_k) \Big\vert S_0 = s, A_0 = a \right ] \nonumber\\
	&&+\mathbf{E}[r(s,a)] \\
	&=& \max_{a\in\mathcal{A}(s)} \mathbf{E}[Q^\ast(S_1,a) \vert S_0 = s, A_0 = a] \nonumber\\
	&& + \mathbf{E}[r(s,a)],
\end{IEEEeqnarray}
where the proof of the last equality can be found in \cite[Section 3.6]{sutton1998introduction}.
The optimal policy $\pi^\ast$ of the agent is
\begin{equation}
\pi^\ast(s) 
= \argmax_{a \in \mathcal{A}(s)} Q^\ast(s,a),
\qquad\forall s\in \mathcal{S}.
\end{equation}
Based on Q-learning in $\epsilon$-greedy policy, agent takes action $a_{i}=\argmax_{a \in \mathcal{A}(s)} Q_i(s_i,a)$ in state $s_{i}\in \mathcal{S}$ with probability $1-\epsilon$ or a random action with probability $\epsilon$ where $\epsilon\in [0,1]$ and $Q_i(s,a)$ is the estimation of the Q function in step i.
In next state $s_{i+1} \in \mathcal{S}$, the agent observes the reward and updates the action-value function $Q_{i+1}(s,a)$ \cite{sutton1998introduction}.

\bibliographystyle{IEEEtran}
\bibliography{IEEEabrv,ref_bib}

\begin{thebibliography}{10}
\providecommand{\url}[1]{#1}
\csname url@samestyle\endcsname
\providecommand{\newblock}{\relax}
\providecommand{\bibinfo}[2]{#2}
\providecommand{\BIBentrySTDinterwordspacing}{\spaceskip=0pt\relax}
\providecommand{\BIBentryALTinterwordstretchfactor}{4}
\providecommand{\BIBentryALTinterwordspacing}{\spaceskip=\fontdimen2\font plus
\BIBentryALTinterwordstretchfactor\fontdimen3\font minus
  \fontdimen4\font\relax}
\providecommand{\BIBforeignlanguage}[2]{{%
\expandafter\ifx\csname l@#1\endcsname\relax
\typeout{** WARNING: IEEEtran.bst: No hyphenation pattern has been}%
\typeout{** loaded for the language `#1'. Using the pattern for}%
\typeout{** the default language instead.}%
\else
\language=\csname l@#1\endcsname
\fi
#2}}
\providecommand{\BIBdecl}{\relax}
\BIBdecl

\bibitem{rappaport2013broadband}
T.~S. Rappaport, F.~Gutierrez, E.~Ben-Dor, J.~N. Murdock, Y.~Qiao, and J.~I.
  Tamir, ``Broadband millimeter-wave propagation measurements and models using
  adaptive-beam antennas for outdoor urban cellular communications,''
  \emph{IEEE Transactions on Antennas and Propagation}, vol.~61, no.~4, pp.
  1850--1859, Apr. 2013.

\bibitem{andrews2016modeling}
M.~R. Akdeniz, Y.~Liu, M.~K. Samimi, S.~Sun, S.~Rangan, T.~S. Rappaport, and
  E.~Erkip, ``Millimeter wave channel modeling and cellular capacity
  evaluation,'' \emph{IEEE Journal on Selected Areas in Communications},
  vol.~32, no.~6, pp. 1164--1179, Jun. 2014.

\bibitem{kutty2016beamforming}
S.~Kutty and D.~Sen, ``Beamforming for millimeter wave communications: An
  inclusive survey,'' \emph{IEEE Communications Surveys \& Tutorials}, vol.~18,
  no.~2, pp. 949--973, 2nd Quart, 2016.

\bibitem{shokri2015millimeter}
H.~Shokri-Ghadikolaei, C.~Fischione, G.~Fodor, P.~Popovski, and M.~Zorzi,
  ``Millimeter wave cellular networks: A {MAC} layer perspective,'' \emph{IEEE
  Transactions on Communications}, vol.~63, no.~10, pp. 3437--3458, Oct. 2015.

\bibitem{UDN}
R.~{Baldemair}, T.~{Irnich}, K.~{Balachandran}, E.~{Dahlman}, G.~{Mildh},
  Y.~{Selen}, S.~{Parkvall}, M.~{Meyer}, and A.~{Osseiran}, ``Ultra-dense
  networks in millimeter-wave frequencies,'' \emph{IEEE Communications
  Magazine}, vol.~53, no.~1, pp. 202--208, Jan. 2015.

\bibitem{rappaport2013millimeter}
T.~S. Rappaport, S.~Sun, R.~Mayzus, H.~Zhao, Y.~Azar, K.~Wang, G.~N. Wong,
  J.~K. Schulz, M.~Samimi, and F.~Gutierrez~Jr, ``Millimeter wave mobile
  communications for {5G} cellular: It will work!'' \emph{IEEE Access}, vol.~1,
  no.~1, pp. 335--349, May 2013.

\bibitem{hanoverHet}
B.~{Yang}, X.~{Yang}, X.~{Ge}, and Q.~{Li}, ``Coverage and handover analysis of
  ultra-dense millimeter-wave networks with control and user plane separation
  architecture,'' \emph{IEEE Access}, vol.~6, pp. 54\,739--54\,750, 2018.

\bibitem{5174147}
``{IEEE} 802.15.3c {P}art 15.3: Wireless medium access control {(MAC)} and
  physical layer {(PHY)} specifications for high rate wireless personal area
  networks {(WPANs)} amendment 2: Millimeter-wave based alternative physical
  layer extension,'' Oct. 2009.

\bibitem{6171799}
``{IEEE} 802.11ad. {P}art 11: Wireless {LAN} medium access control {MAC} and
  physical layer {PHY} specifications - amendment 3: Enhancements for very high
  throughput in the 60 {GH}z band,'' Dec. 2012.

\bibitem{heath2016overview}
R.~W. Heath, N.~Gonzalez-Prelcic, S.~Rangan, W.~Roh, and A.~M. Sayeed, ``An
  overview of signal processing techniques for millimeter wave {MIMO}
  systems,'' \emph{IEEE Journal of Selected Topics in Signal Processing},
  vol.~10, no.~3, pp. 436--453, Apr. 2016.

\bibitem{ghauch2016subspace}
H.~Ghauch, T.~Kim, M.~Bengtsson, and M.~Skoglund, ``Subspace estimation and
  decomposition for large millimeter-wave {MIMO} systems,'' \emph{IEEE Journal
  of Selected Topics in Signal Processing}, vol.~10, no.~3, pp. 528--542, Apr.
  2016.

\bibitem{marzi2016compressive}
Z.~Marzi, D.~Ramasamy, and U.~Madhow, ``Compressive channel estimation and
  tracking for large arrays in mm-wave picocells,'' \emph{IEEE Journal of
  Selected Topics in Signal Processing}, vol.~10, no.~3, pp. 514--527, Apr.
  2016.

\bibitem{Hassanieh}
H.~Hassanieh, O.~Abari, M.~Rodriguez, M.~Abdelghany, D.~Katabi, and P.~Indyk,
  ``Fast millimeter wave beam alignment,'' in \emph{Pro. the Conference of the
  ACM Special Interest Group on Data Communication (ACM SIGCOM)}, 2018, pp.
  432--445.

\bibitem{sur2016beamspy}
S.~Sur, X.~Zhang, P.~Ramanathan, and R.~Chandra, ``Beamspy: Enabling robust 60
  {GHz} links under blockage.'' in \emph{Proc. 13th USENIX Symposium on
  Networked Systems Design and Implementation (USENIX NSDI)}, 2016, pp.
  193--206.

\bibitem{8057188}
A.~Zhou, X.~Zhang, and H.~Ma, ``Beam-forecast: Facilitating mobile 60 {GHz}
  networks via model-driven beam steering,'' in \emph{Pro. IEEE Conference on
  Computer Communications (INFOCOM)}, 2017, pp. 1--9.

\bibitem{MDP}
M.~{Mezzavilla}, S.~{Goyal}, S.~{Panwar}, S.~{Rangan}, and M.~{Zorzi}, ``An
  {MDP} model for optimal handover decisions in mmwave cellular networks,'' in
  \emph{2016 European Conference on Networks and Communications (EuCNC)}, Jun.
  2016, pp. 100--105.

\bibitem{RL2}
Y.~{Sun}, G.~{Feng}, L.~{Zhang}, P.~V. {Klaine}, M.~A. {Iinran}, and
  Y.~{Liang}, ``Distributed learning based handoff mechanism for radio access
  network slicing with data sharing,'' in \emph{IEEE International Conference
  on Communications (ICC)}, May 2019, pp. 1--6.

\bibitem{stevens2008mdp}
E.~Stevens-Navarro, Y.~Lin, and V.~W. Wong, ``An {MDP}-based vertical handoff
  decision algorithm for heterogeneous wireless networks,'' \emph{IEEE
  Transactions on Vehicular Technology}, vol.~57, no.~2, pp. 1243--1254, Mar.
  2008.

\bibitem{zang2018managing}
S.~Zang, W.~Bao, P.~L. Yeoh, B.~Vucetic, and Y.~Li, ``Managing vertical
  handovers in millimeter wave heterogeneous networks,'' \emph{IEEE
  Transactions on Communications}, vol.~67, no.~2, pp. 1629--1644, Feb. 2019.

\bibitem{learninhRL}
Y.~{Sun}, G.~{Feng}, S.~{Qin}, Y.~{Liang}, and T.~P. {Yum}, ``The {SMART}
  handoff policy for millimeter wave heterogeneous cellular networks,''
  \emph{IEEE Transactions on Mobile Computing}, vol.~17, no.~6, pp. 1456--1468,
  Jun. 2018.

\bibitem{side1}
Y.~Koda, K.~Yamamoto, T.~Nishio, and M.~Morikura, ``Reinforcement learning
  based predictive handover for pedestrian-aware mmwave networks,'' in
  \emph{IEEE Conference on Computer Communications Workshops (INFOCOM
  WKSHPS)}.\hskip 1em plus 0.5em minus 0.4em\relax IEEE, 2018, pp. 692--697.

\bibitem{side3learning}
Y.~Koda, K.~Nakashima, K.~Yamamoto, T.~Nishio, and M.~Morikura, ``Handover
  management for mmwave networks with proactive performance prediction using
  camera images and deep reinforcement learning,'' \emph{IEEE Transactions on
  Cognitive Communications and Networking}, vol.~6, no.~2, pp. 802--816, Jun.
  2020.

\bibitem{scaling}
C.~{Fiandrino}, H.~{Assasa}, P.~{Casari}, and J.~{Widmer}, ``Scaling
  millimeter-wave networks to dense deployments and dynamic environments,''
  \emph{Proceedings of the IEEE}, vol. 107, no.~4, pp. 732--745, Apr. 2019.

\bibitem{sur2018towards}
S.~Sur, I.~Pefkianakis, X.~Zhang, and K.-H. Kim, ``Towards scalable and
  ubiquitous millimeter-wave wireless networks,'' in \emph{Proceedings of the
  24th Annual International Conference on Mobile Computing and
  Networking}.\hskip 1em plus 0.5em minus 0.4em\relax ACM, 2018, pp. 257--271.

\bibitem{side2}
R.~Parada and M.~Zorzi, ``Context-aware handover in mmwave {5G} using {UE}'s
  direction of pass,'' in \emph{24th European Wireless Conference}, 2018, pp.
  1--6.

\bibitem{dual2}
V.~{Petrov}, D.~{Solomitckii}, A.~{Samuylov}, M.~A. {Lema}, M.~{Gapeyenko},
  D.~{Moltchanov}, S.~{Andreev}, V.~{Naumov}, K.~{Samouylov}, M.~{Dohler}, and
  Y.~{Koucheryavy}, ``Dynamic multi-connectivity performance in ultra-dense
  urban mmwave deployments,'' \emph{IEEE Journal on Selected Areas in
  Communications}, vol.~35, no.~9, pp. 2038--2055, Sep. 2017.

\bibitem{multicon1}
M.~{Giordani}, M.~{Mezzavilla}, S.~{Rangan}, and M.~{Zorzi}, ``An efficient
  uplink multi-connectivity scheme for {5G} millimeter-wave control plane
  applications,'' \emph{IEEE Transactions on Wireless Communications}, vol.~17,
  no.~10, pp. 6806--6821, Oct 2018.

\bibitem{multicon2}
C.~{Tatino}, I.~{Malanchini}, N.~{Pappas}, and D.~{Yuan}, ``Maximum throughput
  scheduling for multi-connectivity in millimeter-wave networks,'' in
  \emph{16th International Symposium on Modeling and Optimization in Mobile, Ad
  Hoc, and Wireless Networks (WiOpt)}, May 2018, pp. 1--6.

\bibitem{multicon3}
M.~{Gapeyenko}, V.~{Petrov}, D.~{Moltchanov}, M.~R. {Akdeniz}, S.~{Andreev},
  N.~{Himayat}, and Y.~{Koucheryavy}, ``On the degree of multi-connectivity in
  {5G} millimeter-wave cellular urban deployments,'' \emph{IEEE Transactions on
  Vehicular Technology}, vol.~68, no.~2, pp. 1973--1978, Feb. 2019.

\bibitem{hemadeh2017millimeter}
I.~A. Hemadeh, K.~Satyanarayana, M.~El-Hajjar, and L.~Hanzo, ``Millimeter-wave
  communications: Physical channel models, design considerations, antenna
  constructions, and link-budget,'' \emph{IEEE Communications Surveys \&
  Tutorials}, vol.~20, no.~2, pp. 870--913, 2017.

\bibitem{samimi2015probabilistic}
M.~K. Samimi, T.~S. Rappaport, and G.~R. MacCartney, ``Probabilistic
  omnidirectional path loss models for millimeter-wave outdoor
  communications,'' \emph{IEEE Wireless Communications Letters}, vol.~4, no.~4,
  pp. 357--360, 2015.

\bibitem{rappaport2015wideband}
T.~S. Rappaport, G.~R. MacCartney, M.~K. Samimi, and S.~Sun, ``Wideband
  millimeter-wave propagation measurements and channel models for future
  wireless communication system design,'' \emph{IEEE Transactions on
  Communications}, vol.~63, no.~9, pp. 3029--3056, 2015.

\bibitem{sara}
\BIBentryALTinterwordspacing
S.~Khosravi, H.~S. Ghadikolaei, and M.~Petrova, ``Efficient beamforming for
  mobile mm{W}ave networks,'' 2019. [Online]. Available:
  \url{https://arxiv.org/abs/1912.12118v1.}
\BIBentrySTDinterwordspacing

\bibitem{Chong}
E.~Chong and S.~Zak, \emph{An Introduction to Optimization}, ser. Wiley Series
  in Discrete Mathematics and Optimization.\hskip 1em plus 0.5em minus
  0.4em\relax Wiley, 2013.

\bibitem{mobility3}
{Wee-Seng Soh} and H.~S. {Kim}, ``{QoS} provisioning in cellular networks based
  on mobility prediction techniques,'' \emph{IEEE Communications Magazine},
  vol.~41, no.~1, pp. 86--92, Jan. 2003.

\bibitem{mobility2}
A.~N. {Khan} and S.~X. {Jun}, ``A new handoff ordering and reduction scheme
  based on road topology information,'' in \emph{International Conference on
  Wireless Communications, Networking and Mobile Computing}, Sep. 2006, pp.
  1--4.

\bibitem{mobility5}
A.~{Mohamed}, O.~{Onireti}, S.~A. {Hoseinitabatabaei}, M.~{Imran}, A.~{Imran},
  and R.~{Tafazolli}, ``Mobility prediction for handover management in cellular
  networks with control/data separation,'' in \emph{IEEE International
  Conference on Communications (ICC)}, Jun. 2015, pp. 3939--3944.

\bibitem{mobility1}
H.~{Zhang} and L.~{Dai}, ``Mobility prediction: A survey on state-of-the-art
  schemes and future applications,'' \emph{IEEE Access}, vol.~7, pp. 802--822,
  2019.

\bibitem{mobility4}
R.~{Wu}, G.~{Luo}, J.~{Shao}, L.~{Tian}, and C.~{Peng}, ``Location prediction
  on trajectory data: A review,'' \emph{Big Data Mining and Analytics}, vol.~1,
  no.~2, pp. 108--127, Jun. 2018.

\bibitem{mobility11}
Z.~{Abu-Shaban}, X.~{Zhou}, T.~{Abhayapala}, G.~{Seco-Granados}, and
  H.~{Wymeersch}, ``Performance of location and orientation estimation in {5G}
  mmwave systems: Uplink vs downlink,'' in \emph{IEEE Wireless Communications
  and Networking Conference (WCNC)}, Apr. 2018, pp. 1--6.

\bibitem{sutton1998introduction}
R.~Sutton and A.~Barto, \emph{Reinforcement Learning: An Introduction}, ser.
  Adaptive Computation and Machine Learning series.\hskip 1em plus 0.5em minus
  0.4em\relax MIT Press, 2018.

\bibitem{auer2002finite}
P.~Auer, N.~Cesa-Bianchi, and P.~Fischer, ``Finite-time analysis of the
  multiarmed bandit problem,'' \emph{Machine learning}, vol.~47, no. 2-3, pp.
  235--256, 2002.

\bibitem{penetration}
H.~Zhao, R.~Mayzus, S.~Sun, M.~Samimi, J.~K. Schulz, Y.~Azar, K.~Wang, G.~N.
  Wong, F.~Gutierrez, and T.~S. Rappaport, ``28 {GHz} millimeter wave cellular
  communication measurements for reflection and penetration loss in and around
  buildings in {N}ew{Y}ork {City},'' in \emph{Pro. IEEE International
  Conference on Communications (ICC)}, 2013, pp. 5163--5167.

\bibitem{simic2017demo}
L.~Simic, J.~Riihij{\"a}rvi, A.~Venkatesh, and P.~Mahoonen, ``Demo abstract: An
  open source toolchain for planning and visualizing highly directional mm-wave
  cellular networks in the 5{G} era,'' in \emph{IEEE Conference on Computer
  Communications Workshops (INFOCOM WKSHPS)}, 2017, pp. 966--967.

\end{thebibliography}

\end{document}